\documentclass[preprint]{aastex}
\usepackage{graphics,graphicx}
\usepackage[caption=false]{subfig}
\usepackage{subfig}
\usepackage[]{natbib}
\usepackage[]{verbatim}
\usepackage[]{float}
\usepackage{gensymb}
\usepackage{latexsym,amsfonts,amsmath,amssymb}
\newcommand{\upperRomanNumeral}[1]{\uppercase\expandafter{\romannumeral#1}}

\begin{document}
\title{Measurement of Circumstellar Disk Sizes in the Upper Scorpius OB Association with ALMA}

\author{Scott A. Barenfeld}
\affil{California Institute of Technology, Department of Astronomy, MC 249-17, Pasadena, CA 91125}

\author{John M. Carpenter}
\affil{Joint ALMA Observatory, Av. Alonso de C{\'o}rdova 3107, Vitacura, Santiago, Chile}

\author{Anneila I. Sargent}
\affil{California Institute of Technology, Department of Astronomy, MC 249-17, Pasadena, CA 91125}

\author{Andrea Isella}
\affil{Department of Physics and Astronomy, Rice University, 6100 Main St. Houston, TX 77005, USA}

\author{Luca Ricci}
\affil{Department of Physics and Astronomy, Rice University, 6100 Main St. Houston, TX 77005, USA}

\begin{abstract} 

We present detailed modeling of the spatial distributions of gas and dust in 57 circumstellar disks in the Upper Scorpius OB Association observed with ALMA at sub-millimeter wavelengths.  
We fit power-law models to the dust surface density and CO $J$ = 3-2 surface brightness to measure the radial extent of dust and gas in these disks.  We found that 
these disks are extremely compact: the 25 highest signal-to-noise disks have a median dust outer radius of 21 au, assuming an $R^{-1}$ dust surface density profile. Our lack of CO detections in the majority of our sample 
is consistent with these small disk sizes assuming the dust and CO share the same spatial distribution.
Of seven disks in our sample with well-constrained dust and CO radii, four appear to be more 
extended in CO, although this may simply be due to higher optical depth of the CO.  
Comparison of the Upper Sco results with recent analyses of disks in Taurus, Ophiuchus, and Lupus suggests 
that the dust disks in Upper Sco may be $\sim3$ times smaller in size than their younger counterparts, although we 
caution that a more uniform analysis of the data across all regions is needed. We discuss the implications of these results for disk evolution.

\end{abstract}

\section{Introduction}
The past two decades have seen tremendous progress in our understanding of protoplanetary disks \citep[see recent 
reviews by][]{Williams2011,Alexander2014,Dutrey2014,Espaillat2014,Pontoppidan2014,Testi2014,Andrews2015}.  
Submillimeter interferometry has played a crucial role, allowing the gas and dust throughout disks to be studied 
at high spatial resolution. At submillimeter wavelengths, the dust continuum emission from disks is mostly optically thin, making it 
possible to measure dust masses and surface densities. Additionally, a number of molecular species present in disks have 
rotational lines observable in the submillimeter that can be used to study disk temperature, chemistry, kinematics, and mass.

Early submillimeter observations with interferometers focused on young, bright disks, revealing objects which were hundreds of au in size with masses of a few percent of their host stars \citep[e.g.,][]{Kitamura2002,Andrews2007,Isella2007,Andrews2009,Isella2009,Isella2010a,Isella2010b,Guilloteau2011}.
Subsequent observations of fainter disks indicated that smaller sizes and masses may be more typical \citep{Andrews2013,Pietu2014,Testi2016,Hendler2017,Tazzari2017}.  
More recently, a number of studies targeted older protoplanetary disks, which are crucial to our understanding of how disks evolve and dissipate.  Pre-ALMA 
surveys of IC348 \citep{Lee2011}, the Upper Scorpius OB Association 
\citep[hereafter Upper Sco,][]{Mathews2012}, and $\sigma$ Orionis \citep{Williams2013}, revealed a dearth of evolved disks comparable to the brightest objects 
in younger regions, suggesting that older disks are intrinsically fainter than their younger counterparts.

With ALMA, it is possible to conduct large surveys of disks at an unprecedented level of sensitivity, 
revealing the properties of unbiased samples within individual stellar populations.  Thus, \citet{Ansdell2016} 
surveyed the 1-3 Myr old Lupus star-forming region and, from separate measurements of dust and gas masses in 89 disks, 
found evidence that CO is depleted relative to dust compared to interstellar medium (ISM) values \citep[see also][]{Miotello2017}.  \citet{Eisner2016} and \citet{Ansdell2017} 
found evidence for similar depletion in the $<1$ Myr old Orion Nebula cluster and the 3-5 Myr old $\sigma$ Orionis 
region.  In a survey 
of 93 disks in the 2-3 Myr old Chamaeleon star-forming region, \citet{Pascucci2016} found a relationship between disk 
dust mass and stellar mass consistent with other 1-3 Myr old regions, but with a shallower slope than is seen for older disks.  

Crucially, the sensitivity of ALMA enables disk surveys to be extended to more evolved stellar populations \citep[e.g.,][hereafter Paper I]{Carpenter2014,Hardy2015,VanderPlas2016,Barenfeld2016}. 
In particular, Upper Sco provides an ideal target for such studies.  The 5-11 Myr age of this association \citep{Preibisch2002,Pecaut2012} implies that 
its protoplanetary disks are in the last stage of evolution before dissipation \citep{Hernandez2008}.  
Based on ALMA observations of 20 disk-bearing stars in Upper Sco, \citet{Carpenter2014} found tentative evidence that disk dust masses are lower than in 
the younger Taurus star-forming region.  Paper I expanded this sample to include ALMA observations of 106 Upper Sco disks and found that 
the dust masses are on average a factor of 4.5 lower than those in Taurus with high statistical significance.  Of the 58 sources detected in the continuum in this survey, the majority 
were not spatially resolved, implying dust disk radii of a few tens of au or less.  Only 26 sources were detected in CO, suggesting that the CO is 
also confined to a compact emitting area or is heavily depleted relative to the dust.

Here we present a more detailed study of the gas and dust for the disks in the Paper I sample 
and build on our previous results by measuring disk sizes, modeling CO emission, and determining 
the relative distributions of gas and dust.  In Section \ref{sec:samp}, we describe our disk sample and ALMA observations. 
In Section \ref{sec:dust_results}, we detail our methodology for modeling the continuum emission, while our modeling of the CO emission is described in Section \ref{sec:CO_mod}.  We then 
discuss the implications of the gas and dust properties of these disks in Section \ref{sec:discussion}.  Our conclusions are summarized in Section \ref{sec:conclusions}.

\section{Sample \& Observations}
\label{sec:samp}

Our stellar sample for the current work is a subset of the parent sample described in detail in Paper I.  Briefly, our parent sample consisted of 106 stars in Upper Sco 
with spectral types between G2 and M5 (inclusive) and included all 100 candidate G2-M4.75 disk-host 
stars in Upper Sco identified by \citet{Luhman2012} using \emph{Spitzer} and \emph{WISE} observations, as well as six M5 disk-host candidates from \citet{Carpenter2006} 
found using \emph{Spitzer} observations.  

In this work, we analyzed 57 sources detected in Paper I, listed in Tables \ref{tab:cont_fits} and 
\ref{tab:CO_fits}.  Of these sources, 21 were detected in both the 0.88 mm continuum and the CO $J=3-2$ line 
at 345.79599 GHz, 34 were only detected in the continuum, and two were detected only in CO.  
Five of the sources analyzed are classified as debris/evolved transitional disks by \citet{Luhman2012}.  
We consider the remaining sources to be primordial disks (see Paper I).

Three extremely bright continuum sources were identified in Paper I, 2MASS J15583692-2257153, 2MASS J16042165-2130284, 
and 2MASS J16113134-1838259, which have continuum flux densities of $174.92\pm0.27$ mJy, $218.76\pm0.81$ mJy, and $903.56\pm0.85$ mJy at 0.88mm, respectively.  
2MASS J15583692-2257153 exhibits an azimuthal asymmetry in the continuum, while 2MASS J16042165-2130284 shows the large inner cavity of a transitional disk.  2MASS J16113134-1838259 
is more than 20 times brighter in the continuum than any of the sources we are including in this paper and exhibits possible disk winds and tidal interactions with a stellar companion \citep{Salyk2014}.  Since these 
systems are not representative of typical disks in Upper Sco, we excluded them from the present analysis and focused instead on understanding the broader population of ordinary disks.  
\citet{Zhang2014} presented a detailed analysis of 2MASS J16042165-2130284 \citep[see also][]{Mayama2012,vanDerMarel2015,Pinilla2015,Dong2017}.

The ALMA observations were obtained in Cycle 0 and Cycle 2 using the 12 m array (see Paper I). All observations used band 7, with a mean frequency 
of 340.7 GHz for Cycle 0 and 341.1 GHz for Cycle 2 (0.88 mm) and a total bandwidth of 7.5 GHz. One spectral window was configured with channel 
widths of 0.488 MHz (0.429 km s$^{-1}$, the spectral resolution is twice the channel width) to observe the CO $J$ = 3-2 line at 345.79599 GHz.  The 
observations had angular resolution between 0$\overset{''}{.}$35 and 0$\overset{''}{.}$73 with a median of 0$\overset{''}{.}$37 and a continuum rms ranging from 0.13 mJy/beam to 0.26 mJy/beam, with a median of 0.15 mJy/beam.

\section{Continuum Modeling}
\label{sec:dust_results}

Our goal in modeling the continuum data was to determine the radial extent of the dust for the 55 continuum-detected disks in our sample. 
To accomplish this, we compared our observed visibilities to the synthetic visibilities of a model disk, deriving the model visibilities from an assumed 
dust density distribution in a self-consistent way.  

For our model disk, we parameterized the dust surface density as a function of radius using a truncated power law: 
\begin{equation}
\Sigma(R) = \Sigma_0\left(\frac{R}{10 \mathrm{au}}\right)^{-1},
\end{equation}
for $R$ between the dust inner and outer radii, $R_{in}$ and $R_{dust}$, with $\Sigma = 0$ outside of this range.  We fixed 
$R_{in}$ at the values found by the SED fitting of \citet{Mathews2013} for 
the 24 sources we share with their survey.  For our remaining sources, we set $R_{in}$ to be equal to the dust sublimation radius, calculated based on the stellar luminosities from Paper I.  The choice 
of $R_{in}$ does not impact our results.  

The continuum signal-to-noise ratio for the majority of our sample is too low to simultaneously constrain the dust outer radius and the slope of the surface density power law.  
We therefore adopted a fixed $R^{-1}$ parameterization for the surface density, which is fairly typical for disks 
\citep[e.g.,][]{Kitamura2002,Andrews2007,Isella2007,Isella2010a,Guilloteau2011}. 
Assuming $\Sigma \propto R^{-0.5}$ and $\Sigma \propto R^{-1.5}$ power laws resulted
in slightly smaller or larger disk sizes, respectively, but did not qualitatively impact our results or conclusions.

Some authors \citep[e.g.,][]{Hughes2008,Andrews2009,Isella2009, Andrews2010a, Andrews2010b, Hughes2010, Isella2010a, Isella2010b, Andrews2011,Guilloteau2011,Andrews2012} have parameterized the disk surface 
density using the self-similar solution for a viscously evolving disk, 
which can be approximated as a power law with an exponential tail.  However, given the evolutionary state of the Upper Sco disks, it is not clear that this description is appropriate.  
Other effects, such as the inward radial migration of dust grains \citep[e.g.,][]{Weidenschilling1977}, may also change the surface density profile.  \citet{Birnstiel2014} modeled dust transport 
in a viscously evolving disk and found that grain migration results in a dust surface density well-described by a power law with a sharp outer edge.  We therefore adopted a power-law surface density 
for our analysis.  Broken power-law models have also been used to model dust surface density \citep[e.g.,][]{Hogerheijde2016} but given the low signal-to-noise of the majority of our detections, we opted 
for a single power-law model with fewer free parameters.

The vertical disk structure was parameterized using the commonly assumed Gaussian vertical density structure of an isothermal disk \citep[e.g.,][]{Isella2007}:
\begin{equation}
\rho(R,z) = \frac{\Sigma(R)}{h\sqrt{2\pi}}\exp\left(\frac{-z^2}{2h^2}\right), 
\end{equation}
The disk scale height, $h$, was assumed to be a power law function of radius:
\begin{equation}
h=h_{0}\left(\frac{R}{100\mathrm{au}}\right)^p.
\end{equation}
We allowed $h_{0}$ and $p$ to vary with priors based on the SED fits of 
\citet{Mathews2013}.  Our choice of these priors 
does not affect our conclusions.  Thus, if a disk was included in the \citet{Mathews2013} sample, we used their best fit 
values for $h_{0}$ and $p$ with uncertainties of $1$ au and 0.05 to create normally distributed priors.  If a disk was not in the \citet{Mathews2013} sample, we 
assumed a normally distributed prior for $p$, with a mean of 1.13 (the 
median value of $p$ for their sample of 45 disks) and standard deviation of 0.05.  For $h_{0}$, 
three quarters of the \citet{Mathews2013} disks had $h_{0}<10$ au, so we assumed uniform priors from 0-10 au and 10-20 au, with the probability of $h_{0}>10$ au equal to 
one-third of the probability of $h_{0}<10$ au. 

We also assumed a constant dust opacity throughout the disk. Previous multi-wavelength studies of disks at millimeter and centimeter wavelengths suggested radial 
variations in dust opacity due to grain growth \citep{Isella2010a,Banzatti2011,Guilloteau2011,Perez2012,Perez2015,Trotta2013,Menu2014,ALMA2015,Guidi2016,Tazzari2016}. Similar 
radial variations may be present in our Upper Sco disks, and, in fact, are predicted by models of dust transport and evolution \citep[e.g.,][]{Dullemond2005,Birnstiel2010}.  
However, there is no way to constrain the dust opacity based on our single-wavelength data.
Thus, we used a uniform dust opacity, calculated as a function of wavelength using Mie theory for dust grains composed of a mix of carbons, ices, and silicates \citep[e.g.,][]{Pollack1994}, with a grain size distribution of 
$n(a)\propto a^{-3.5}$ and a maximum grain size of 1 cm.  Only $\Sigma_0$, which is inversely proportional to opacity for a given 
flux density, was sensitive to our choice of maximum grain size.  

The Monte Carlo radiative transfer code RADMC-3D \citep{Dullemond2012} was used to determine the temperature throughout the model dust disk due to stellar irradiation.  
We adopted the stellar parameters derived in Paper \upperRomanNumeral{1}.
We assumed a minimum temperature of 10 K at any location in the disk to account for other heat sources 
such as radioactive decay and cosmic rays \citep[e.g.,][]{Dallessio2001,Woitke2015}.
RADMC-3D was then used to generate an image of the model disk for a given inclination and position angle. The Fourier transform of this image provided a grid of model visibilities, which was interpolated at our observational $uv$ points.  We used the $\chi^2$ difference 
between the model and observed visibilities (real and imaginary) to calculate the likelihood of the current set of model parameters:
\begin{equation}
L = \sum\exp\left[-\frac{(\mathrm{Re}_{mod}-\mathrm{Re}_{obs})^2}{2\sigma_{vis}^2}-\frac{(\mathrm{Im}_{mod}-\mathrm{Im}_{obs})^2}{2\sigma_{vis}^2}\right],
\end{equation}
where $\frac{1}{\sigma_{vis}^2}$ is the visibility weight.  The observed visibilities were corrected to center the disk at the phase center of observations using the disk positions determined in Paper I.
The Markov chain Monte Carlo implementation \emph{emcee} \citep{Foreman-Mackey2013}  
was used to constrain $\Sigma_0$, $R_{dust}$, $h_0$, $p$, inclination, and position angle.

Table \ref{tab:cont_fits} contains the most likely values for $\Sigma_0$, $R_{dust}$, $h_{0}$, $p$, inclination, and position angle from the continuum fitting, together with their 
uncertainties.
The values of each parameter sampled by the MCMC in the fitting of each source gave the final probability distribution of that parameter.  The reported values in Table \ref{tab:cont_fits} 
were taken from the peak of these distributions.  The uncertainties were defined as the bounds of the range around the peak containing 68.3\% of the integrated probability.  
The dust disks range from 4 to 173 au in radius, although 82\% of the disks have radii less than 50 au.  Figure \ref{fig:dust_results} 
shows an image of the best-fit model of each source, along with the observed image and residuals.  This figure also shows the deprojected observed and best-fit model visibilities 
as a function of baseline length.  

Five sources, 2MASS J16032225-2413111, 2MASS J16054540-2023088, 2MASS J16111330-2019029, 2MASS J16123916-1859284, 
and 2MASS J16135434-2320342, exhibited $5\sigma$ emission in the residual images of their best-fit models.  We fit point source models to the residual visibilities of these sources; their 
continuum flux densities and positions relative to the primary disks are listed in Table \ref{tab:secondaries}.  
The NASA/IPAC Extragalactic Database lists no known background galaxies at the positions of these secondary sources. 
2MASS J16054540-2023088 was identified as a double line spectroscopic binary by \citet{Dahm2012}, while none 
of the other sources with secondary emission are known binaries.

The secondary sources of 2MASS J16032225-2413111, 2MASS J16123916-1859284, 
and 2MASS J16135434-2320342 were also identified in Paper I.  
The secondary sources of 2MASS J16054540-2023088 and 2MASS J16111330-2019029 were too close to their respective primary sources to have been identified in Paper I, 
but can be seen in Figure \ref{fig:dust_results} as non-axisymmetric extended emission on the eastern side of the 
primaries.  The secondary source to 2MASS J16135434-2320342, by far the brightest in our sample, is clearly visible as a second disk to the east of the primary.  We used the MCMC 
fitting method described above to determine the dust properties of this source after subtracting our best-fit model for the primary disk from the observed visibilities and shifting the phase center 
to the secondary disk.  The best-fit parameters are given in Table \ref{tab:cont_fits} for both components of 2MASS J16135434-2320342.

\section{CO Modeling}
\label{sec:CO_mod}

\subsection{CO Surface Brightness Fitting}
\label{sec:CO_fitting}

Our modeling approach for the CO data was similar to that for the continuum.  We used the continuum-subtracted visibilities 
for each of the 23 CO detections to measure the radial extent of gas.  Due to the likelihood of optically thick CO 
emission, we fit the CO surface brightness of the disks instead of a physical surface density.  For each source, we used the velocity 
range corresponding 
to the $J=3-2$ emission line, as determined by Paper \upperRomanNumeral{1}, to generate integrated ``moment 0 visibilities.''  
We then fit to the real and imaginary part of these visibilities as described in Section \ref{sec:dust_results}.    
For consistency with our continuum fitting, we assumed an azimuthally symmetric disk with surface brightness described by a truncated power law,
\begin{equation}
S(R) = S_0\left(\frac{R}{10 \mathrm{au}}\right)^{-\gamma}
\end{equation}
for $R<=R_{CO}$ and $S=0$ beyond $R_{CO}$.
We used \emph{emcee} to fit for the surface brightness normalization, $S_0$, the power law slope, $\gamma$, the outer radius, $R_{CO}$, the inclination, and position angle.

Table \ref{tab:CO_fits} presents the best-fit parameters of our CO model fitting.  Best-fit values and uncertainties were defined as described in Section \ref{sec:dust_results} for the dust modeling. 
Observed, model, and residual images for the CO are shown in Figure \ref{fig:CO_results}, along with the deprojected observed and best-fit visibilities of each source.
We found CO outer radii ranging from 6 to 430 au.  2MASS J16154416-1921171 was the largest CO disk in our sample, with a radius of 430 au. 
Examination of the CO channel maps suggested contamination by a surrounding molecular cloud and we therefore excluded this source from further CO analysis. With this 
source excluded, our largest CO outer radius is 169 au.
It is worth noting that the gaseous disks may extend beyond our measured CO outer radii; CO will be subject to freeze-out and photodissociation in the outer parts of disks \citep[though some CO will return to 
the gas phase through non-thermal desorption, e.g.,][]{Oberg2016} while H$_2$ and other gaseous molecules can survive out to these regions.

\subsection{Expected CO Fluxes}
\label{sec:CO_depletion}

Only 21 of the 55 continuum-detected sources in our sample were also detected in CO.  The relatively low number 
of CO detections suggested that the CO is either heavily depleted relative to the dust or has a compact emitting 
area due to small disk sizes.  In this section, we test the latter possibility.  We used the results of our continuum modeling to predict the CO $J=3-2$ line flux from our continuum-detected disks assuming the gas and dust share the same spatial distribution.  

To estimate the expected CO $J=3-2$ line fluxes for our continuum detected disks, we used the posterior distributions of $\Sigma_0$, $R_{dust}$, $h_0$, $p$, inclination, and 
position angle from our MCMC continuum fits to generate a sample of model dust disks for each source. We then added CO to these disks by assuming that the CO and dust share the same temperature structure and spatial distribution with 
a gas-to-dust ratio of 100 and a CO to H$_2$ ratio of $7\times10^{-5}$ by number \citep[][and references therein]{Beckwith1993,Dutrey1996}.
If a source was detected in CO, we sampled CO outer radii ($R_{CO}$) from the posterior distribution of our surface brightness 
fitting and extended the model CO disk out to these radii.  If $R_{CO}$ is larger than $R_{dust}$, we used our 3$\sigma$ upper limits on the total continuum flux between 
$R_{dust}$ and $R_{CO}$ to calculate an upper limit on the dust mass in this annulus, assuming optically thin emission and a dust temperature of 10 K.  This dust mass upper 
limit was then converted into a uniform dust surface density between $R_{dust}$ and $R_{CO}$, and the disk was populated with CO as described above.  

We took into account the removal of CO from the gas phase by freeze-out and photodissociation.  
At any location in a model disk where the temperature was less than 20 K, we assumed the CO was frozen onto dust grains and had 
an abundance of zero \citep{Collings2003,Bisschop2006}. While a small fraction of this CO will re-enter the gas phase through UV photodesorption \citep{Oberg2009a,Oberg2009b,Fayolle2011,Chen2014} and cosmic ray heating 
\citep{Hasegawa1993}, modeling of these processes has shown the effects on CO observations to be negligible 
\citep{Oberg2015}.
A common method for treating photodissociation in disks is to assume a minimum column density of H$_2$ that will shield CO from destruction by stellar and 
interstellar Ultraviolet and X-Ray radiation.  \citet{Visser2009} modeled a molecular cloud exposed to the interstellar radiation field and found 
that an H$_2$ column density of $10^{21}$ g cm$^{-2}$ would shield CO.  Detailed modeling \citep{Aikawa2006,Gorti2008} and observations \citep{Qi2011} of 
photodissociation in disks around accreting young stars found similar results.  Thus, it is often assumed that CO in disks will only survive below a 
vertical column density of $10^{21}$ g cm$^{-2}$ of H$_2$ \citep{Williams2014,Walsh2016}.

In the young circumstellar disks modeled in this way, high energy radiation produced by stellar accretion 
dominated over the interstellar radiation field \citep{Zadelhoff2003,Visser2009}, providing an abundant source of UV and X-ray photons.  
This may not be the case for the more evolved disks in Upper Sco, however.  \citet{Dahm2009} found that only 7 out of 
a sample of 35 disk-bearing Upper Sco sources showed signs of accretion, and that the median accretion 
rate of these 7 sources was an order of magnitude lower than that of younger disks in Taurus.  Therefore, the disks in the present sample are 
likely to be exposed to much weaker radiation fields than younger, more strongly accreting disks, and will require less material to shield CO.  To reflect this uncertainty in the minimum 
shielding column density required, we treated photodissociation in two ways.  First, we followed the typical assumption for younger disks and assumed that if the vertical column density of 
H$_2$ above any location in the disk was below $10^{21}$ cm$^{-2}$, CO would be photodissociated to a density of zero.  As an alternative, we also calculated the expected CO flux without any 
photodissociation, which we consider a conservative upper limit on the amount of gaseous CO that survives and therefore on the model CO flux.  We note that once enough CO survives to 
become optically thick, the shielding column density and precise amount that survives has little impact on the expected flux.  Ignoring photodissociation entirely is an approximation of this scenario.

\begin{figure}[!h]
\centering
\subfloat{\includegraphics[width=0.45\linewidth]{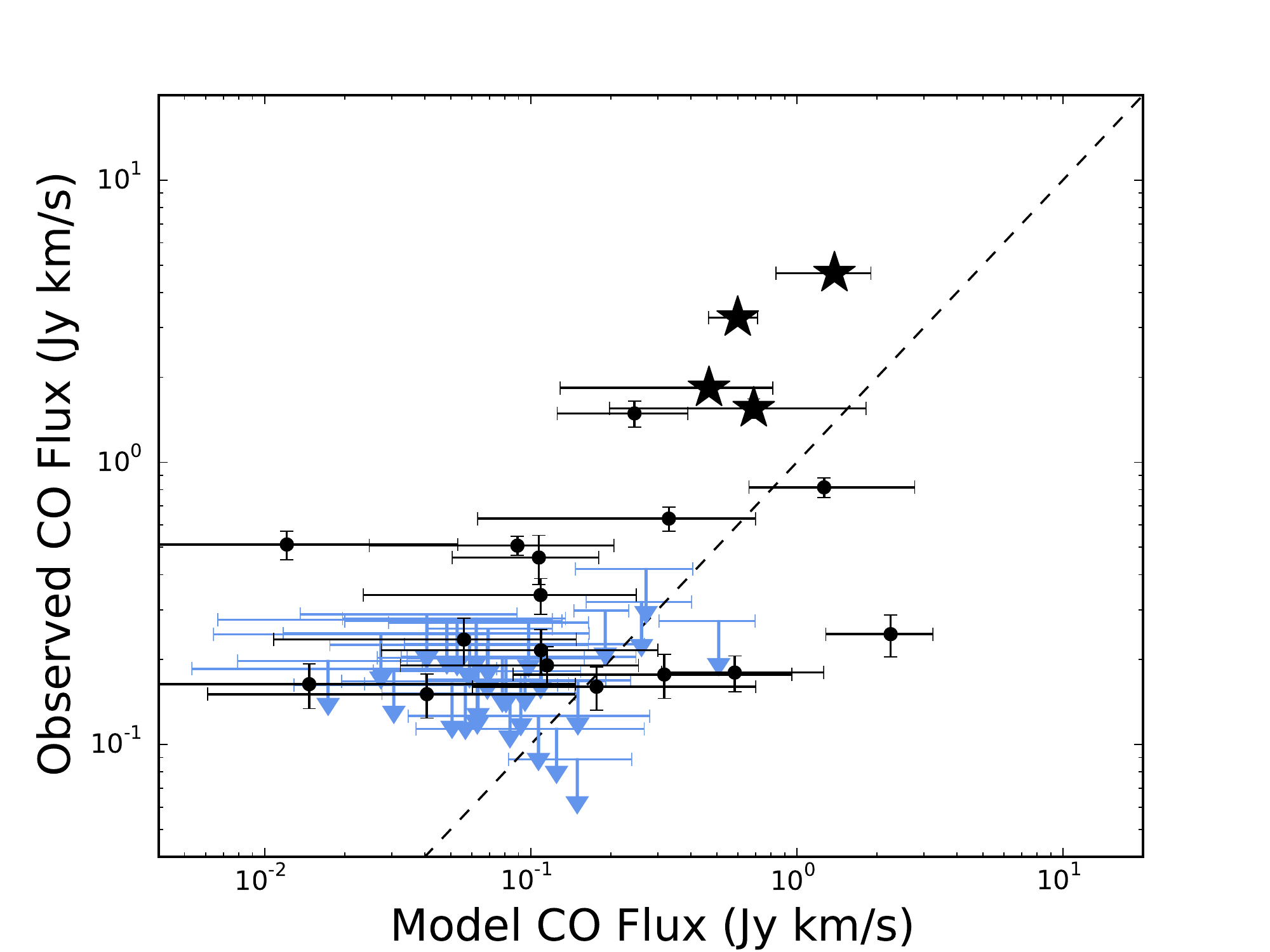}}
\qquad
\subfloat{\includegraphics[width=0.45\linewidth]{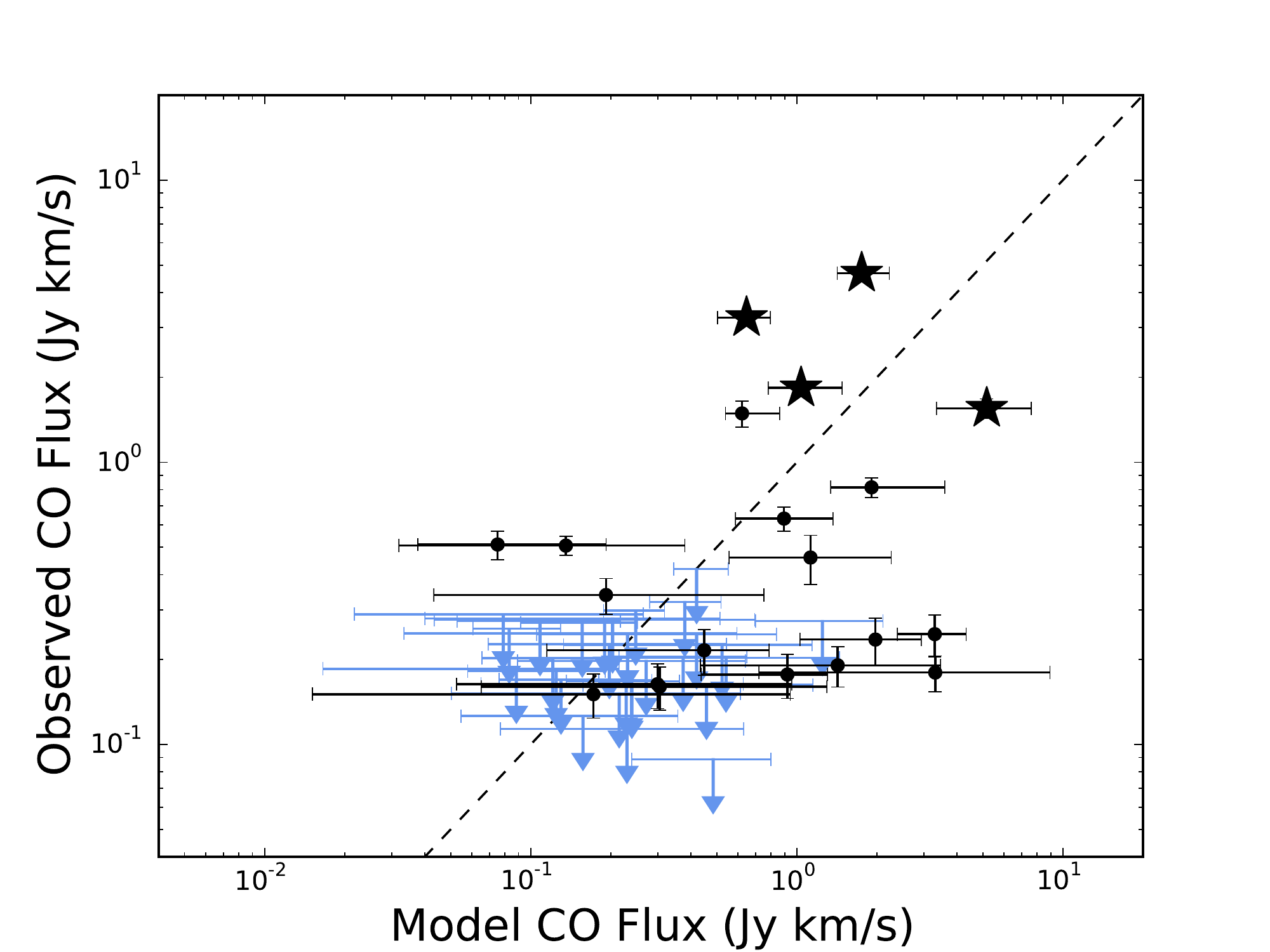}}
\caption{Observational and model fluxes for CO $J=3-2$ emission for sources detected in the continuum, for models with (left panel) and without (right panel) photodissociation. Model fluxes were calculated based on the expected CO emission given the disk dust properties, as discussed in Section \ref{sec:CO_depletion}. CO detections are shown as black points, while upper limits are shown with blue arrows. The four well-constrained sources with larger CO outer radii than dust outer radii are shown as stars.  Horizontal error bars represent the 68.3\% confidence range for the model fluxes. The dashed line represents agreement between the model and observed fluxes.
}
\label{fig:fluxcomp}
\end{figure}

With these model CO disks, we used RADMC3D to calculate the CO $J=3-2$ flux over the velocity range determined for each source in Paper I. We repeated this process for every continuum-detected source in our sample, generating a distribution of 
model CO fluxes for each. Figure \ref{fig:fluxcomp} shows a comparison of our model CO fluxes with our observed fluxes from Paper I for the two treatments of 
photodissociation described above. Among the sources detected in CO, there 
was considerable scatter in the observed fluxes relative to the model fluxes.  This reflects the uncertainties of our modeling procedure, both statistical from our uncertain 
dust model parameters and systematic relating to our assumptions regarding gas temperature and gas-to-dust ratio, as well as our treatments of freeze-out and photodissociation. 
Without photodissociation, CO fluxes increased by as much as an order of magnitude.  This was due to cases where the combination of freeze-out and photodissociation truncated 
the CO disk inside of the observed $R_{CO}$, reducing the emitting area of the disk.
For the sources not detected in CO, the model fluxes were consistent with observational upper limits.

\section{Discussion}
\label{sec:discussion}

\subsection{Dust Disk Sizes}
\label{sec:DDS}
\begin{figure}[!h]
\centerline{\includegraphics[scale=1.0]{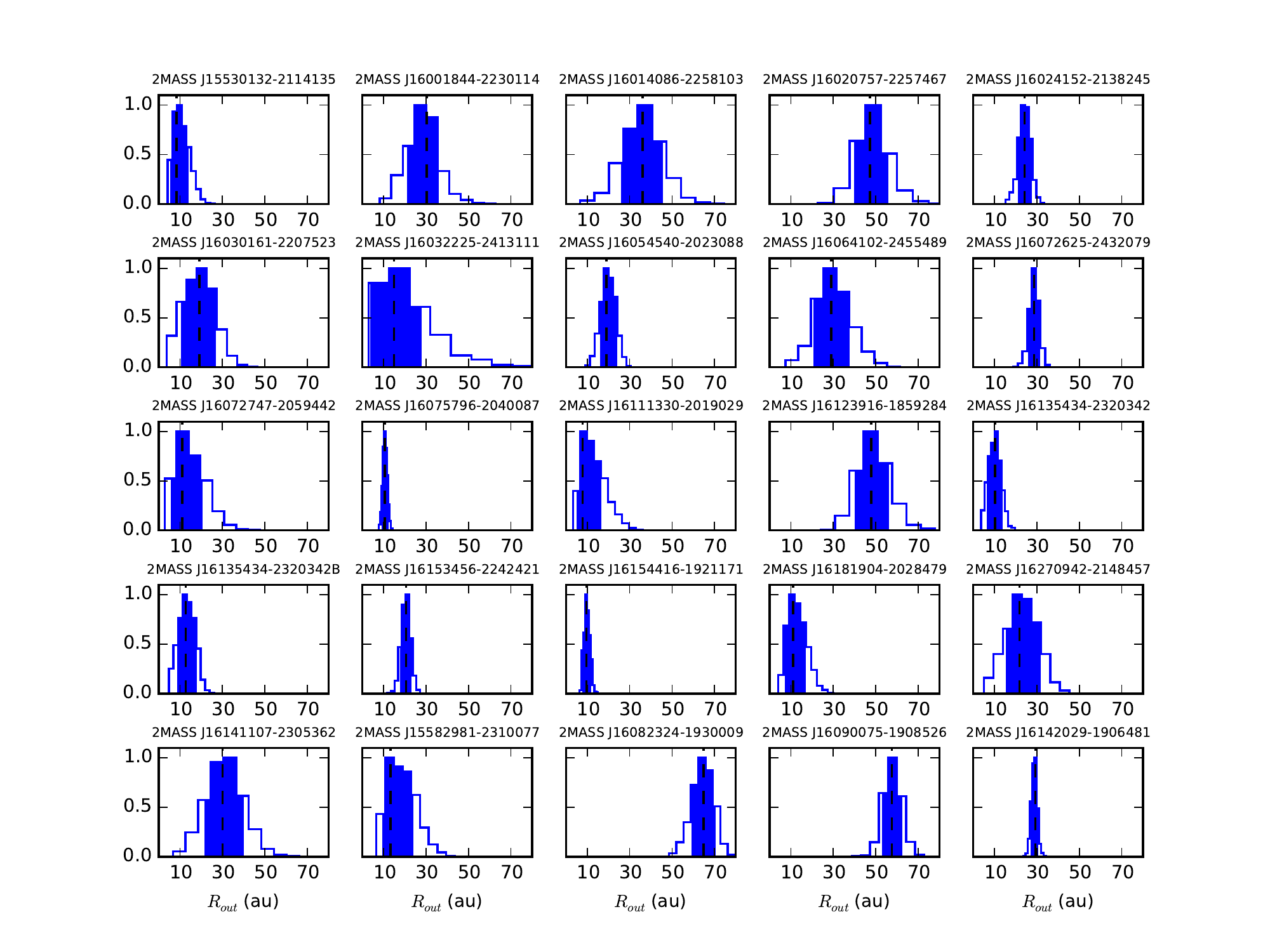}}
\caption{Posterior distributions of dust outer radius for the 25 disks with continuum signal-to-noise of at least 15. The distributions are sharply peaked around the best-fit values (dashed lines), indicating that 
these disks are well-constrained to be compact. The blue shaded regions show the 68.3\% confidence range for the outer radii.
}
\label{fig:radii_posteriors}
\end{figure}

Based on our derived dust outer radii, the majority of the continuum-detected dust disks in our sample are quite compact.  Empirically, we determined that to constrain the dust outer radius ($R_{dust}$)
to better than a factor of 2 required a signal-to-noise of at least 15.  The 25 disks that meet this threshold have dust outer radii ranging from 8 to 65 au, with a median of 21 au.  
Only two disks, 2MASS J16082324-1930009 and 2MASS J16090007-1908526, have radii larger than 50 au. Note that this 
excludes 2MASS J15583692-2257153, 2MASS J16042165-2130284, and 2MASS J16113134-1838259 (see Section \ref{sec:samp}), 
all of which appear to be larger than 65 au in radius based on visual inspection of their continuum images (see Paper I).  
Figure \ref{fig:radii_posteriors} shows the posterior probability distributions of the outer radius for the 25 high signal-to-noise disks. The distributions are sharply-peaked around the best-fit 
value, with no significant probability tails extending out to larger radii. Thus, while we cannot rule out that the 
dust surface density follows a different distribution than $R^{-1}$, such as a power law with a different slope or with an exponential tail, any such distribution must fall off rapidly at or near our best fit outer 
radii.

While we lack a sample of younger disks analyzed in the same way to compare with our 5-11 Myr old Upper Sco sample, we do see evidence that the dust disks in Upper Sco are more compact than 
younger disks.  \citet{Tripathi2017} measured the sizes of 50 disks primarily located in the 1-2 Myr old Taurus and Ophiuchus star-forming regions by fitting ``Nuker'' profiles \citep{Lauer1995} to 
the continuum emission. More than half of the stars in this sample have spectral types earlier than K9, compared to only 2 of 25 stars in our high signal-to-noise sample, J16141107-2305362 and 2MASS J16154416-1921171. We therefore include only spectral types 
K9-M5 when comparing the present Upper Sco sample to these younger disks.  Pre-main sequence stars of these spectral types are fully convective and evolve at approximately constant temperature \citep[e.g.,][]{Siess2000}, 
making spectral type a good proxy for stellar mass even when comparing stars of different ages.
For this spectral type range, \citet{Tripathi2017} found effective radii, defined as the radius containing 68\% of the disk continuum flux, ranging from 19 au to 182 au, with a median of 48 au.  
Assuming optically thin 
dust emission and constant midplane dust temperature in the outer regions of our Upper Sco disks, where most of the dust mass resides, we can define an effective radius for our sample as containing 68\% 
of the total dust mass, which will be approximately equivalent to the radius containing 68\% of the continuum flux. With this definition, the effective radii of our high signal-to-noise sources with spectral types 
K9-M5 range from 5 au to 44 au, with a 
median of 14 au. These effective radii may in fact be overestimated for the Upper Sco disks, as the innermost region of the disks will have higher dust temperatures than the outer regions, causing the disk 
continuum flux to be slightly more concentrated at small radii than the dust mass. However, since we found that the disks in Upper Sco appear to be smaller than the younger disks of \citet{Tripathi2017}, this effect 
strengthens our conclusions. 

In a separate study of young disks, \citet{Tazzari2017} fit for the outer radii of 22 disks in the 1-3 Myr old Lupus star-forming region. The authors used a power-law with an exponential cutoff to parameterize the dust surface density, defining the effective radius as that which contains 95\% of the dust mass. We again exclude 2MASS J16141107-2305362 and 2MASS J16154416-1921171, restricting our comparison to stars between 0.15 and 0.7 M$_{\odot}$. For this 
stellar mass range, \citet{Tazzari2017} measure effective radii ranging from 18 au to 129 au, with a median of 55 au. Calculating the radii of our disks containing 95\% 
of the dust mass, we find a range of 7 au to 62 au, with a median of 20 au.  Taken at face-value, these results suggest that the disks in Upper Sco are smaller than those found in Taurus, Ophiuchus, and Lupus by a factor of 
$\sim$3. However, we caution however that a self-consistent analysis of all these disks needs to be performed to confirm this trend.

Finally, we note that \citet{Hendler2017} measured dust outer radii from the spectral energy distributions of 
11 young disks around very low mass stars and brown dwarfs in the Taurus and Chamaeleon I star forming regions, finding disk sizes similar to those we see 
in Upper Sco. This younger sample probes lower stellar masses than the present Upper Sco sample, and is therefore not 
directly comparable. However, \citet{VanderPlas2016} used ALMA to image the disks around 7 very low mass stars and 
brown dwarfs in Upper Sco. None of these objects were spatially resolved, constraining them to also be compact ($
\lesssim40$ au). Follow-up studies of these low mass stellar and substellar systems can be used to determine 
if the reduction in dust disk sizes with age observed here extends to lower stellar masses.

\subsection{Comparing Dust and CO}
\label{sec:dust_gas_comp}
\begin{figure}[!h]
\centerline{\includegraphics[scale=1.0]{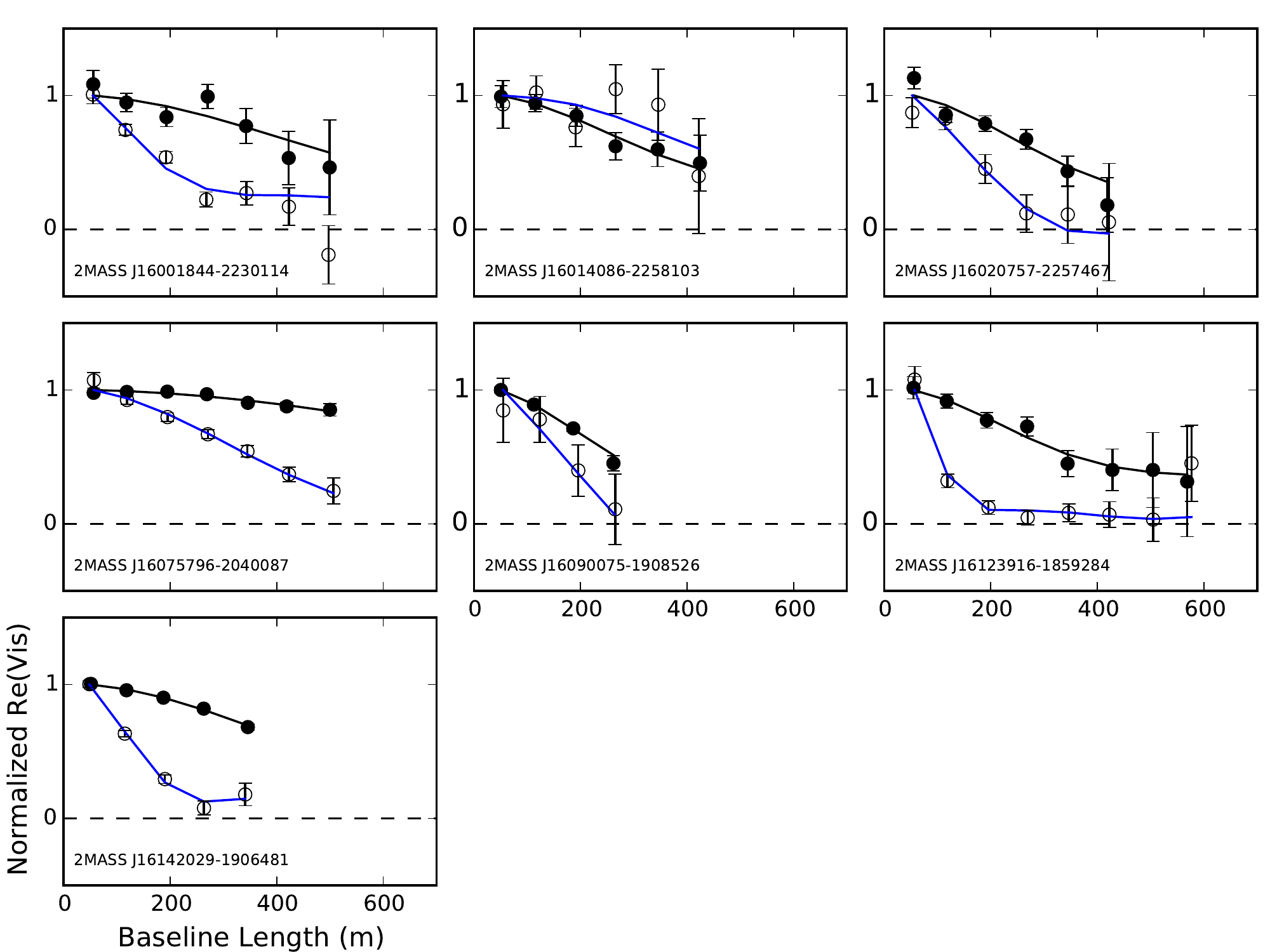}}
\caption{Continuum (black circles) and CO (open circles) deprojected visibilities for the sources with well-constrained dust and CO outer radii.  The black and blue curves show the best-fit models 
for the dust and CO, respectively.  Four sources, 2MASS J16001844-2230114, 2MASS J16075796-2040087, 2MASS J16123916-1859284, and 2MASS J16142029-1906481, exhibited detectable CO emission extending 
beyond their dust emission.}
\label{fig:radii_vis}
\end{figure}

\begin{figure}[!h]
\centerline{\includegraphics[scale=1.0]{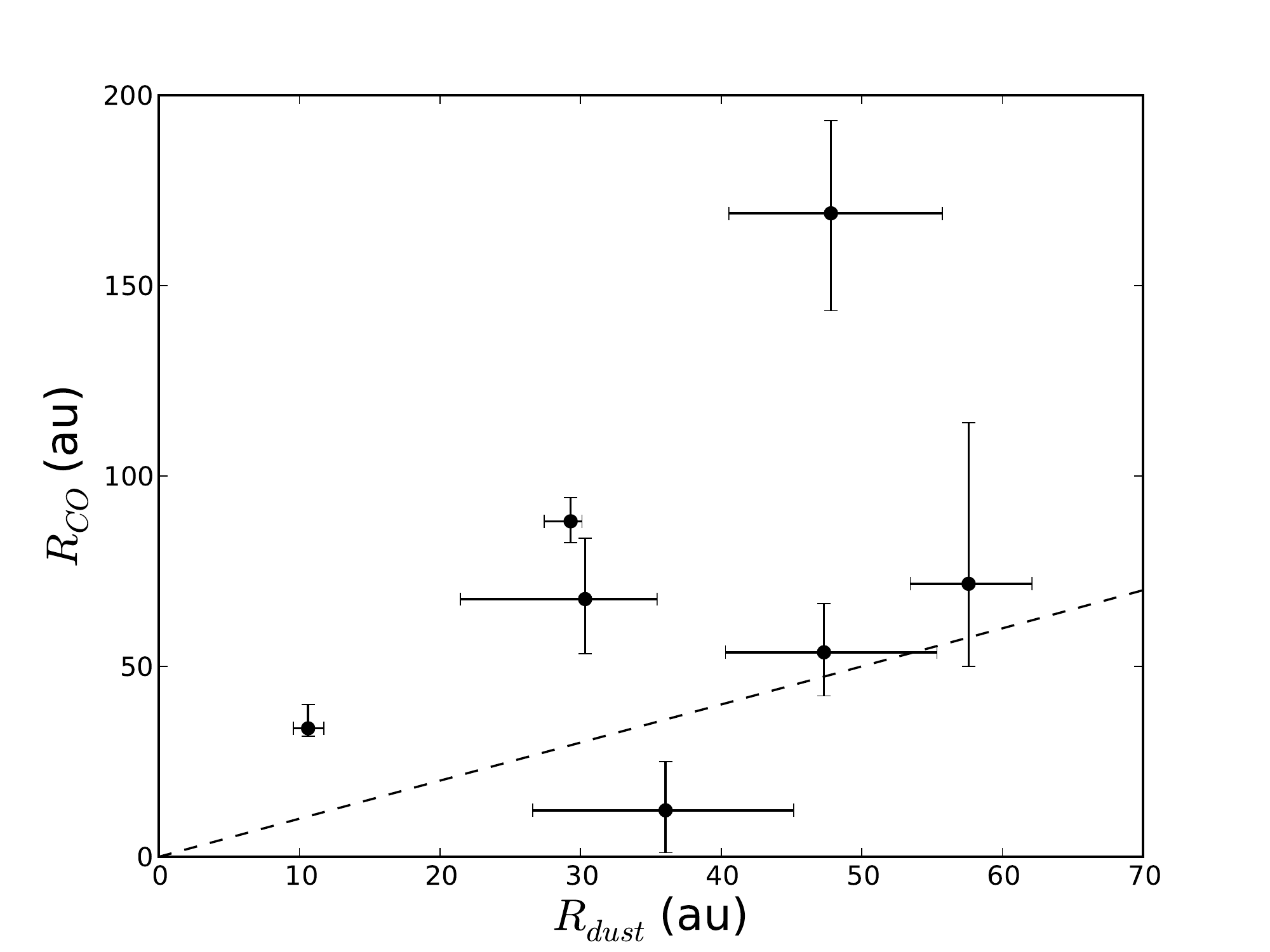}}
\caption{Best fit CO and dust outer radii for sources where both are well constrained. Four sources, 2MASS J16001844-2230114, 2MASS J16075796-2040087, 2MASS J16123916-1859284, and 2MASS J16142029-1906481, 
have CO outer radii larger than their dust outer radii.}
\label{fig:radii_comp}
\end{figure}

Empirically, we found that to measure $R_{CO}$ to better than a factor of 2 required a 
CO signal-to-noise of at least 8, as measured from the moment 0 maps in Paper I.  This threshold is lower than that of continuum data 
due to the CO model fitting having one less free parameter than the continuum fitting. Also, the CO emission tends to be more extended than the 
continuum emission, allowing for a smaller fractional uncertainty on the outer radius for a given signal-to-noise.
For the 9 disks with well-constrained CO outer radii we measured radii ranging from 12 to 169 au, with a median of 54 au, 
excluding 2MASS J16154416-1921171 (see Section \ref{sec:CO_fitting}).  Only 3 of these 9 sources 
have CO radii less than 50 au.  Figure \ref{fig:radii_vis} displays the continuum and CO deprojected visibilities, with their best-fit models, 
for the 7 sources with well-constrained dust and CO outer radii.  Figure \ref{fig:radii_comp} shows the outer radii for these sources.  Four sources, 2MASS J16001844-2230114, 2MASS J16075796-2040087, 2MASS J16123916-1859284, 
and 2MASS J16142029-1906481, have detectable CO emission extending to larger radii than the detectable dust emission. 
Previous observations of younger disks also revealed CO emission extending beyond any detectable continuum emission \citep[e.g.,][]{Pietu2005,Isella2007,Panic2009,Andrews2012} and enhanced gas-to-dust 
ratios at large radii \citep{Isella2016}. 

However, optical depth effects must be taken into account \citep{Hughes2008,Facchini2017}.  
A low surface density 
tail of dust may extend beyond the apparent dust outer radius, with its optically thin emission undetected. Emission from CO, on the other hand, is optically thick down to low surface densities, and therefore is more likely to be detected in the outer parts of a disk even if the dust emission is weak.  
To test this possibility for the four sources with CO potentially extending beyond the dust, we used the predicted CO fluxes of Section \ref{sec:CO_depletion} and 
Figure \ref{fig:fluxcomp}, where these four sources are shown as stars.  These models assumed a low surface density tail of dust, consistent with 
our upper limits, between the apparent dust and CO outer radii, with standard CO abundances relative to the dust.  We found that the predicted CO line fluxes of these sources were consistent with the observed fluxes, although this 
was dependent on the assumed photodissociation prescription. Therefore, while these disks may in fact have an enhanced gas-to-dust ratio in the outer disk due 
to inward grain migration \citep[e.g.,][]{Birnstiel2014}, we could not rule out a standard gas-to-dust ratio, with a drop in surface density of both gas and dust beyond 
the apparent dust outer radius.  Further observations that place deeper limits on the dust surface density in the outer disk and/or include the optically thin 
isotopologues of CO to estimate gas surface densities can be used to distinguish between these two cases.

Previous studies of 1-3 Myr old disks have found evidence for a low CO abundance relative to dust throughout the disk
\citep{Dutrey2003,Chapillon2008,Williams2014,Hardy2015,Long2017}. 
\citet{Ansdell2016} used observations of optically thin $^{13}$CO and C$^{18}$O emission to measure the gas-to-dust 
ratios of 62 disks in 1-3 Myr old Lupus complex, finding that the majority of sources had a gas-to-dust ratio 
below the ISM value of 100 assuming an ISM abundance of CO relative to H$_2$ \citep[see also][]{Miotello2017}. 
\citet{Ansdell2017} found similar results for the 3-5 Myr old $\sigma$ Orionis region.  To the extent that CO traces 
the total gas mass, this has important implications for disk evolution and the relative timescales of gas and dust 
dissipation in disks.  However, chemical processing of the gas in disks is expected to lower the CO to H$_2$ ratio \citep{Kama2016,Yu2017}. 
Gas mass measurements using the HD 112 $\mu$m 
line showed that CO in disks may be depleted relative to hydrogen by up to two orders of magnitude \citep{Bergin2013,McClure2016}.

If the processes causing these low CO abundances continue to the age of Upper Sco, this could explain 
the lack of CO detections in over half of our continuum-detected disks.  However, as our analysis in Section \ref{sec:CO_depletion} shows, 
CO depletion is not required to explain these non-detections, given the signal-to-noise of the data.  The small sizes of these disks alone are sufficient to explain the lack 
of detectable CO emission.  For the sources where 
we do see CO emission, we could not constrain the total mass in CO due to the likelihood that the emission is optically thick. 
Upper Sco represents a crucial data point to study the relative evolution of gas and dust in disks, but to do so requires additional 
sensitivity and/or observations of $^{13}$CO and C$^{18}$O to constrain the gaseous CO mass.

\subsection{Implications for Disk Evolution}

The evolved nature of the disks in our Upper Sco sample presents an opportunity to use the properties of these disks to improve our 
understanding of disk evolution. Paper I and \citet{Ansdell2016} showed that the dust disks in Upper Sco are on average a factor of 3-4.5 
less massive than those in Taurus and Lupus \citep[see also][]{Pascucci2016}. Taken at face-value, the indication of dust disks being more compact as well in Upper Sco (Section \ref{sec:DDS}) 
implies that at least some of this mass is lost through the disappearance of millimeter grains in the outer disk. These grains may be completely removed 
from the millimeter grain population of the system, either through photoevaporation of the outer disk \citep[e.g.,][]{Owen2012,Alexander2014,Gorti2015} or 
through growth into larger bodies \citep[][and references therein]{Testi2014}. On the other hand, inward migration of grains from the outer disk \citep[e.g.,][]{Weidenschilling1977,Birnstiel2014} 
may cause the inner disk to become optically thick, in effect hiding the dust mass of the outer disk and making the disk appear to be less massive.

We tested these scenarios by comparing the dust surface densities of the Upper Sco disks in this work and of the Lupus disks measured by \citet{Tazzari2017} to determine if the amount of inner disk dust has increased by the age 
of Upper Sco. We used the best-fit surface density 
normalizations of both studies, $\Sigma_0$, representing the surface density at the normalization radius of 10 au, as a proxy for inner disk surface density. The $\Sigma_0$ values 
of \citet{Tazzari2017} are measured assuming a dust opacity at $890\mu$m of 3.37 cm$^2$ g$^{-1}$ so we scaled their surface densities to match 
our assumed dust opacity of $4.94$ cm$^{2}$ g$^{-1}$.  In addition, the authors report the inferred \emph{gas} surface density at 10 au assuming a gas-to-dust ratio of 100. 
We therefore divided by their assumed gas-to-dust ratio to recover the dust surface density.  Applying these corrections and restricting our comparison to stars between 0.15 and 0.7 M$_{\odot}$ 
as in Section \ref{sec:DDS}, we found $\Sigma_0$ values 
ranging from $4.9\times10^{-3}$ g cm$^{-2}$ to $0.71$ g cm$^{-2}$ for Lupus and 
$6.2\times10^{-3}$ g cm$^{-2}$ to $0.23$ g cm$^{-2}$ for the high signal-to-noise Upper Sco 
disks. The mean of $\log\Sigma_0$ is $-1.04$ for Lupus, with a standard deviation of $0.64$, while Upper Sco has a mean $\log\Sigma_0$ of $-1.65$, with a standard deviation of $0.42$. In addition, the mean $\log\Sigma_0$ value we 
find for Upper Sco implies an inner disk that only becomes optically thick inside of $\sim1$ au, assuming 
$\Sigma\propto R^{-1}$. We therefore see no evidence of inner disks increasing in
dust surface density between Lupus and Upper Sco.

However, our modeling assumes that the dust in these disks is distributed smoothly in radius following a simple parameterization, which may not be the case. \citet{Tripathi2017} suggested that disks may be composed of 
optically thick substructure with a filling factor of a few tens of percent to explain an observed correlation between 
disk size and luminosity in 1-2 Myr old disks in Taurus and Ophiuchus. This idea is supported by 
theoretical models of gas and dust interactions in disks, which predict that small-scale gas pressure 
maxima can trap dust grains and create concentrations of 
optically thick continuum emission \citep[e.g.,][]{Whipple1972,Pinilla2012}. Recent high resolution observations with ALMA have revealed a number of disks 
exhibiting such substructure \citep{ALMA2015,Andrews2016,Cieza2016,Isella2016,Perez2016,Zhang2016,Loomis2017}. 
If young disks such as those in Lupus and Taurus are in general composed of optically thick substructures with filling 
factors less than 1, appearing optically thin to lower resolution observations, this could provide a way to hide dust grains
migrating from the outer disk and cause the disk to appear less massive as it decreases in size. As long as the filling factor 
of the substructure does not increase, lower resolution observations such as those presented here and in \citet{Tazzari2017} would 
not detect the increase in dust surface density.

\begin{figure}[!h]
\centerline{\includegraphics[scale=1.0]{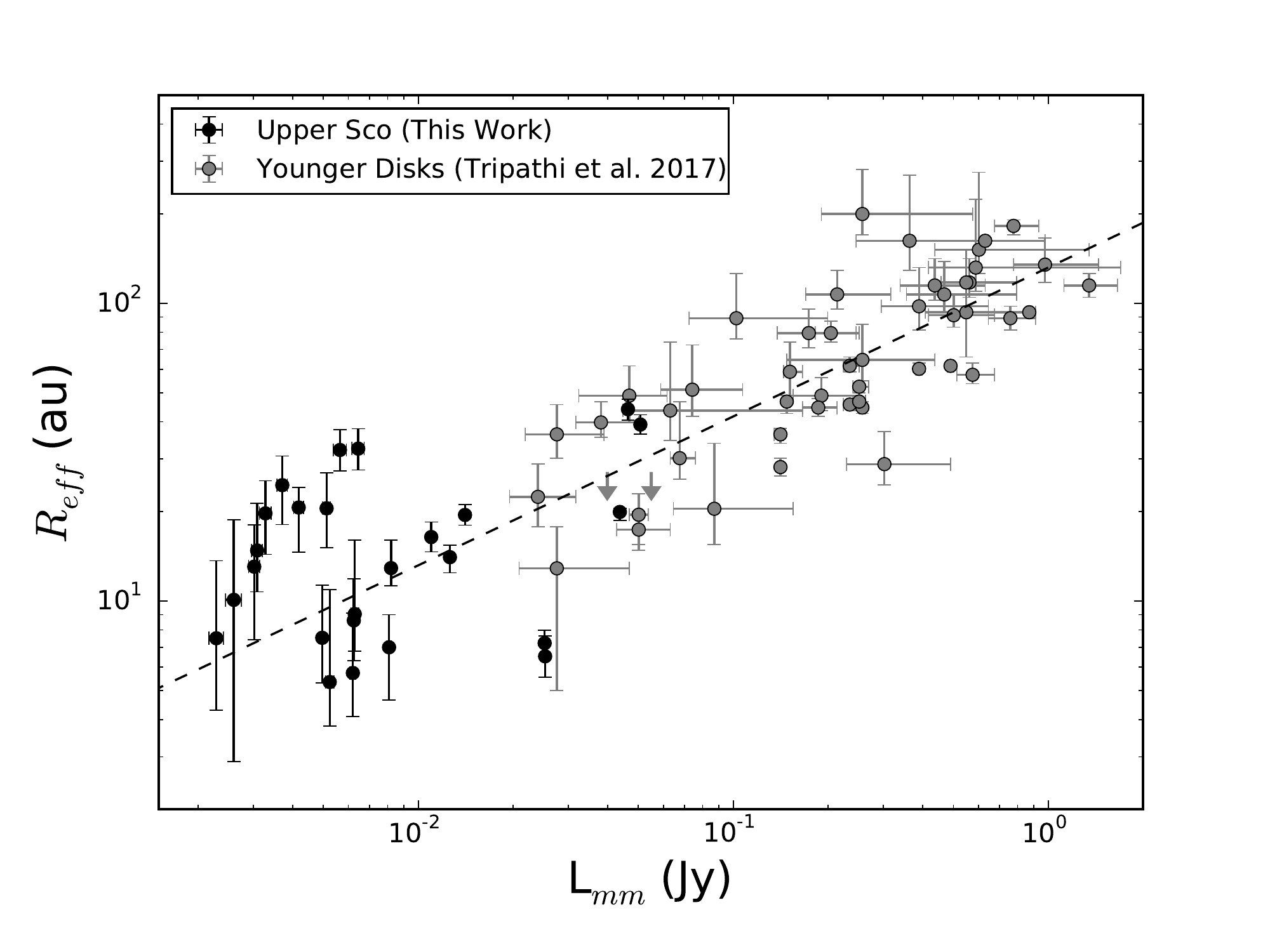}}
\caption{Dust effective radius versus continuum flux density at 0.88 mm for circumstellar disks in Taurus and Ophiuchus \citep[gray points,][]{Tripathi2017} and Upper Sco (black points, this work).  Flux densities have been 
scaled to a distance of 140 pc. The dashed line shows the disk size-luminosity relation from \citet{Tripathi2017}, which has been extrapolated to the flux densities of the Upper Sco disks. The Upper Sco sources shown are those 
with a continuum signal-to-noise of at least 15.
}
\label{fig:disk_size_disk_flux}
\end{figure}

Figure \ref{fig:disk_size_disk_flux} compares the effective radii ($R_{eff}$) and continuum fluxes scaled to 140 pc ($L_{mm}$) for Upper Sco and the young stars analyzed by \citet{Tripathi2017}. 
The \citet{Tripathi2017} disk size-luminosity relation, $R_{eff}\propto L_{mm}^{0.50\pm0.07}$, is also shown, 
extrapolated to the scaled flux densities of Upper Sco.
Despite the difference in age and luminosity, our disks lie approximately along this extrapolation. 
We therefore conclude that our data is qualitatively consistent with the \citet{Tripathi2017} disk size-luminosity 
relation. Thus, if the disk size-luminosity correlation is caused by optically thick substructures, the filling factor of these substructures will be roughly the same in Upper Sco as in Taurus and Ophiuchus. Higher resolution observations of samples of disks of different ages can help to further constrain the evolution of inner disks, allowing the dust surface density and optical depth to be more precisely measured as a function 
of radius and possibly detecting substructure.

\section{Summary}
\label{sec:conclusions}

We have presented detailed modeling of 57 circumstellar disks in the Upper Scorpius OB Association observed with ALMA. Our sample excludes the three 
brightest continuum disks observed in Paper I, 2MASS J15583692-2257153, 2MASS J16042165-2130284, 
and 2MASS J16113134-1838259, instead focusing on more typical Upper Sco disks.  Power-law model fits of the 
dust surface density to the continuum observations yielded the radial extent of dust in these disks. Similar model fits to the CO surface brightness 
of the disks measured the extent of CO. Using our modeling results, we compared the spatial extents of dust and CO Upper Sco disks and calculated a range of expected 
CO fluxes, comparing these model fluxes to our observed values from Paper I. 
The key conclusions of this paper are as follows:
\begin{enumerate}

\item Of the 25 analyzed disks with a continuum signal-to-noise of at least 15, we find a median dust outer radius of 21 au. Only two of these disks had dust outer radii larger than 50 au, with none greater than 65 au, 
assuming an $R^{-1}$ power-law dust surface density. While this excludes the three brightest continuum sources in our sample, which appear to be more extended, 
it is clear that the majority of the high signal-to-noise dust disks in Upper Sco are extremely compact.

\item Among our seven disks with well-constrained dust and CO outer radii, four exhibited CO radii significantly larger than their dust radii.  Given the signal-to-noise of the continuum and CO data, this may simply be a result 
of higher optical depths of the CO line. More sensitive observations, especially of $^{13}$CO and 
C$^{18}$O, are needed to determine whether there is a true deficit of dust in the outer regions of these disks.

\item Assuming that the CO and dust share the same spatial distribution, the lack of CO detections 
in most of the disks is consistent with the small disk sizes inferred from the continuum.

\item Dust disks in Upper Sco are a factor of $\sim3$ more compact than those in Taurus, Ophiuchus, and Lupus. Assuming that the continuum emission is optically thin, the lower disk masses in Upper Sco relative to Taurus and Lupus \citep[Paper I;][]{Ansdell2016} appear to be primarily due to removal of material in the outer disk. We caution 
however that a more uniform analysis between samples is needed.

\item The disks in Upper Sco fall along the same size-luminosity correlation found by \citet{Tripathi2017}. If the origin of this correlation is caused by the presence of optically thick substructures, the filling factor of such 
structures is similar between Upper Sco and the young disks studied by \citet{Tripathi2017}.

\end{enumerate}

\newpage
\acknowledgements

We thank the referee and statistics editor for their useful comments, which improved this manuscript.
We are grateful to the ALMA staff for their assistance in the data
reduction. The National Radio Astronomy Observatory is a facility of the
National Science Foundation operated under cooperative agreement by Associated
Universities, Inc. This paper makes use of the following ALMA data:
ADS/JAO.ALMA\#2011.0.00966.S and ADS/JAO.ALMA\#2013.1.00395.S. ALMA is a partnership of ESO (representing its
member states), NSF (USA) and NINS (Japan), together with NRC (Canada) and NSC
and ASIAA (Taiwan), in cooperation with the Republic of Chile. The Joint ALMA
Observatory is operated by ESO, AUI/NRAO, and NAOJ.  This material is based upon work supported by the 
National Science Foundation Graduate Research Fellowship under Grant No. DGE‐1144469.  S.A.B. acknowledges support 
from the NSF Grant No. AST-1140063.  J.M.C. acknowledges 
support from the National Aeronautics and Space Administration under Grant No. 15XRP15\_20140 issued through 
the Exoplanets Research Program. A.I. acknowledges support from the NSF Grant No. AST-1535809 and the
National Aeronautics and Space Administration Grant
No. NNX15AB06G. This publication makes use
of data products from the Two Micron All Sky Survey, which is a joint project
of the University of Massachusetts and the Infrared Processing and Analysis
Center/California Institute of Technology, funded by the National Aeronautics
and Space Administration and the National Science Foundation. This publication 
makes use of data products from the \emph{Wide-field Infrared Survey Explorer}, which 
is a joint project of the University of California, Los Angeles, and the Jet 
Propulsion Laboratory/California Institute of Technology, funded by the National 
Aeronautics and Space Administration.  This research has made use of the NASA/IPAC 
Extragalactic Database (NED) which is operated by the Jet Propulsion Laboratory, 
California Institute of Technology, under contract with the National Aeronautics and 
Space Administration. This work is based [in part] on observations 
made with the Spitzer Space Telescope, which is operated by the Jet Propulsion Laboratory, 
California Institute of Technology under a contract with NASA.


\begin{figure}[!h]
\centering

\subfloat[][]{
\includegraphics[width=\textwidth]{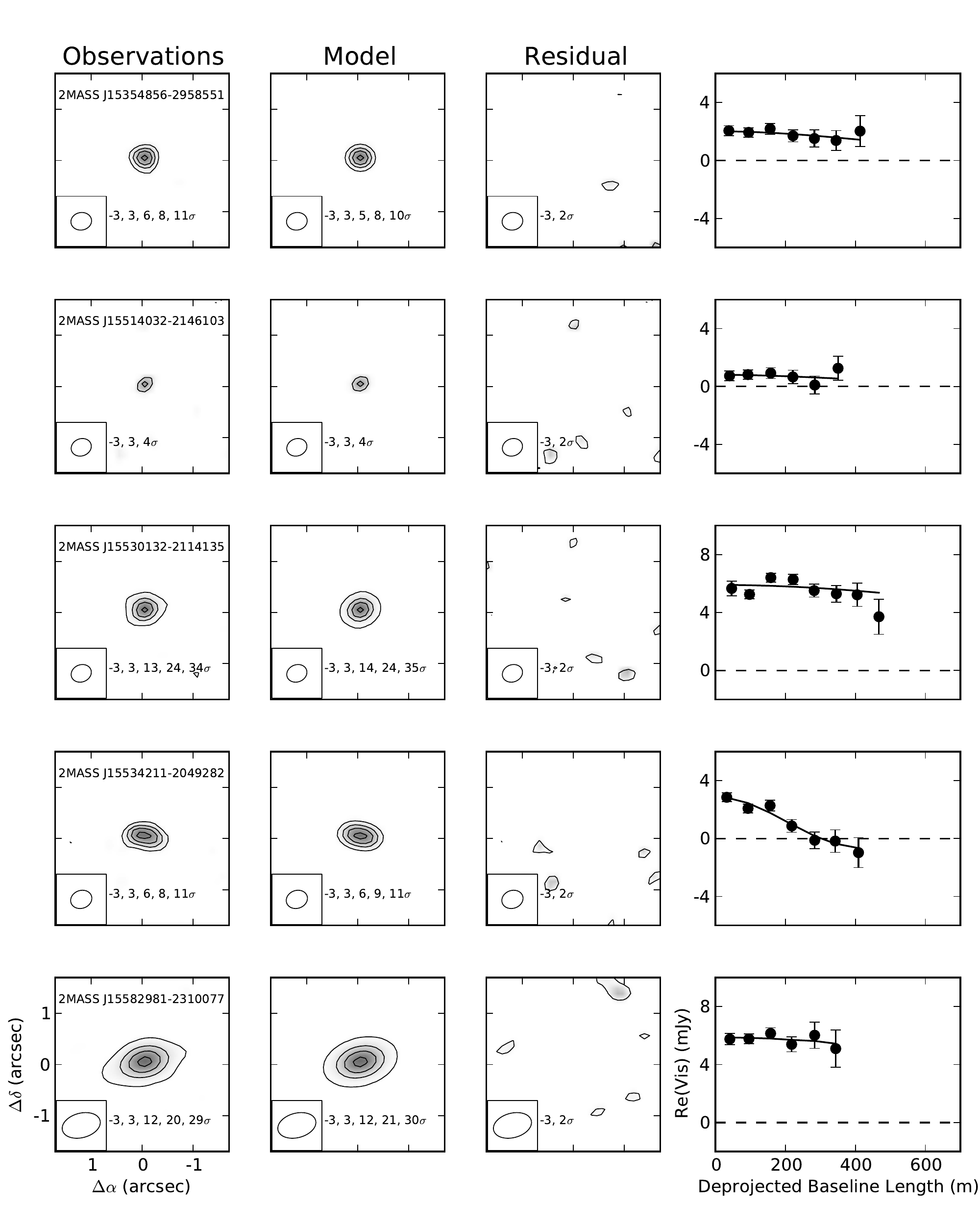}
}
\caption{ALMA 0.88 mm observed, model, and residual images corresponding to the best fit dust model parameters for each source.  The real part of the deprojected visibilities for the observations 
(solid points) and best fit model (solid curve) are also shown as a function of baseline length.}
\label{fig:dust_results}
\end{figure}

\begin{figure}
\ContinuedFloat
\centering
\subfloat[][]{
\includegraphics[width=\textwidth]{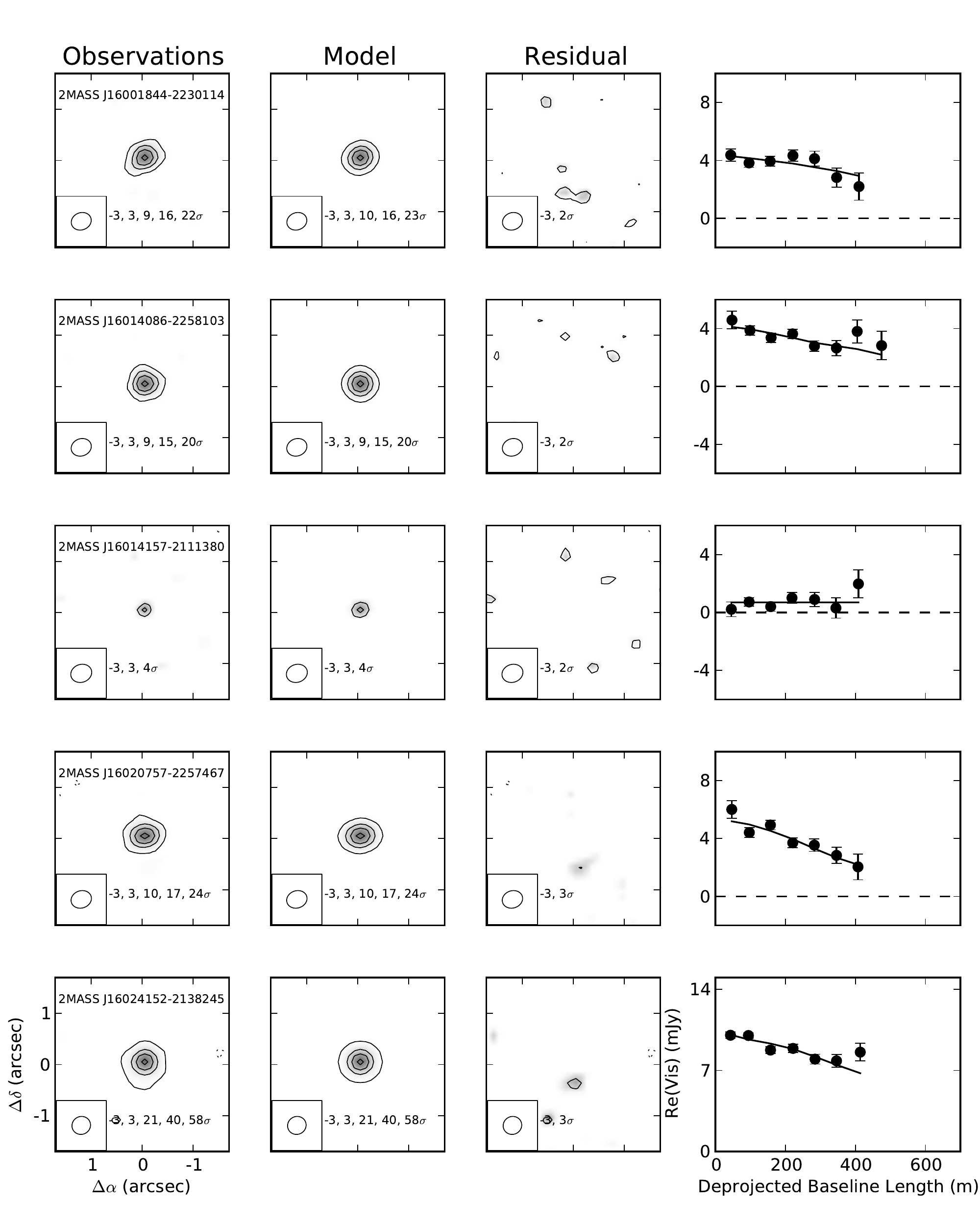}
}
\caption{Continued.}
\label{fig:dust_results}

\end{figure}

\begin{figure}
\ContinuedFloat
\centering
\subfloat[][]{
\includegraphics[width=\textwidth]{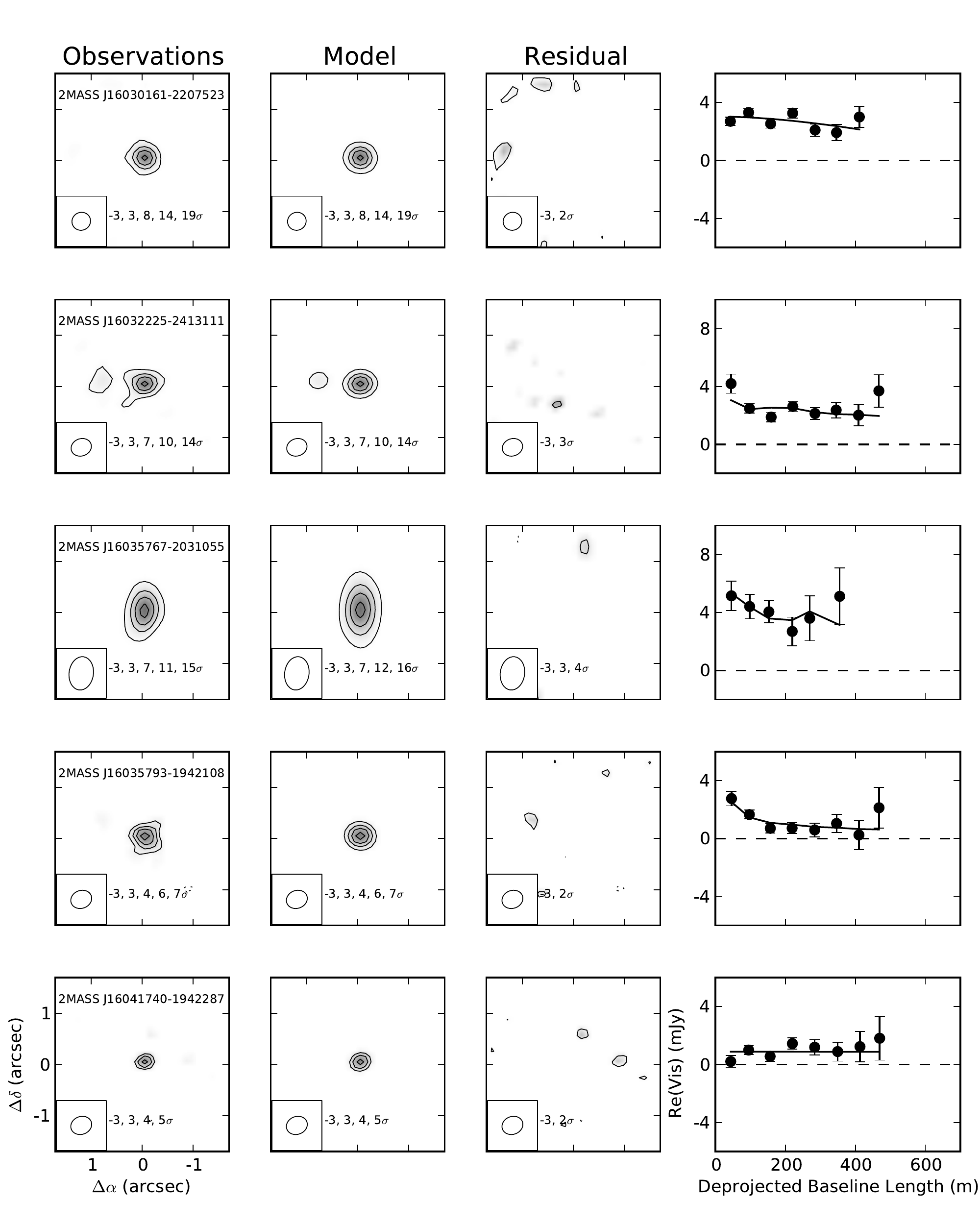}
}
\caption{Continued}
\label{fig:dust_results}
\end{figure}

\begin{figure}
\ContinuedFloat
\centering
\subfloat[][]{
\includegraphics[width=\textwidth]{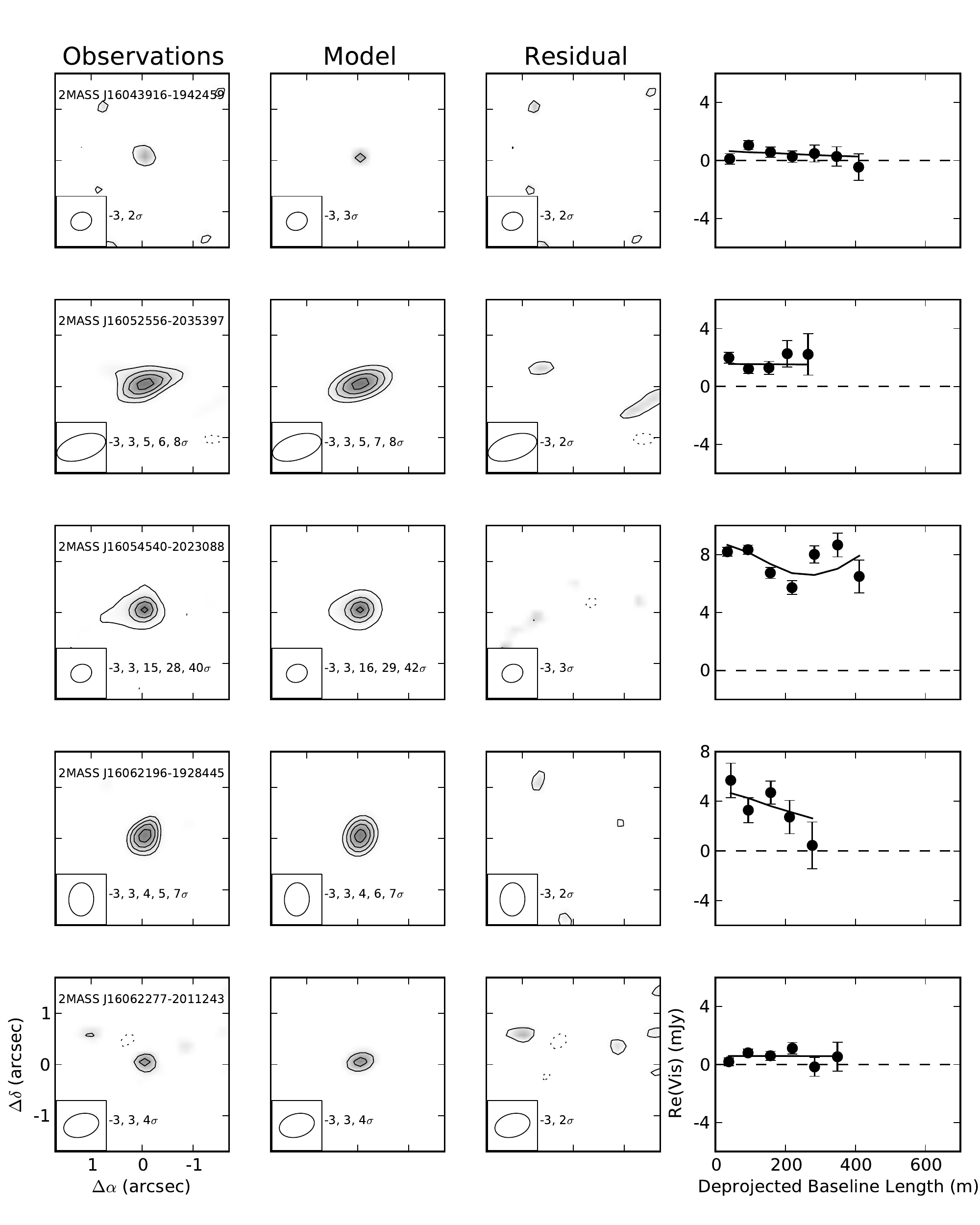}
}
\caption{Continued}
\label{fig:dust_results}
\end{figure}

\begin{figure}
\ContinuedFloat
\centering
\subfloat[][]{
\includegraphics[width=\textwidth]{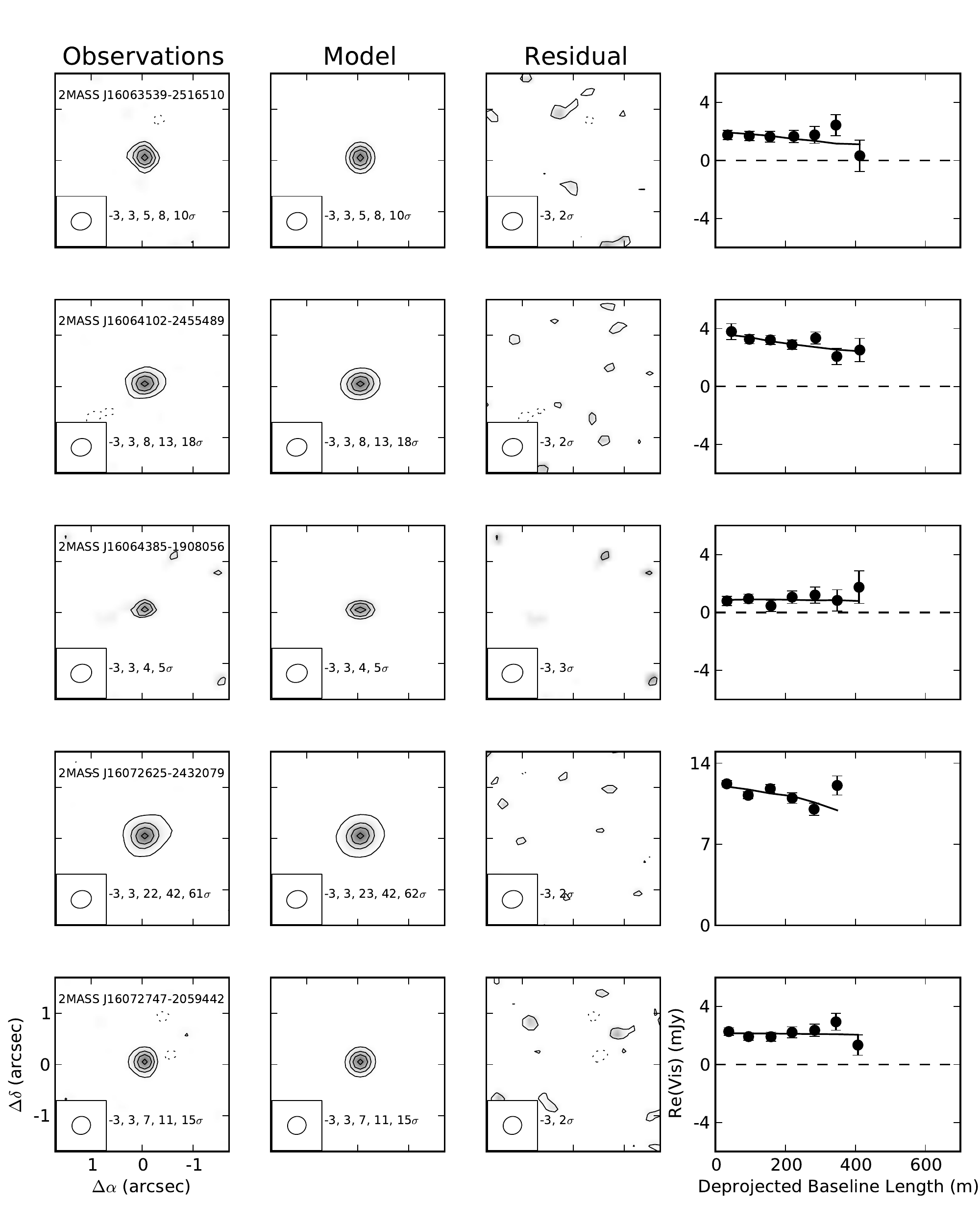}
}
\caption{Continued}
\label{fig:dust_results}
\end{figure}

\begin{figure}
\ContinuedFloat
\centering
\subfloat[][]{
\includegraphics[width=\textwidth]{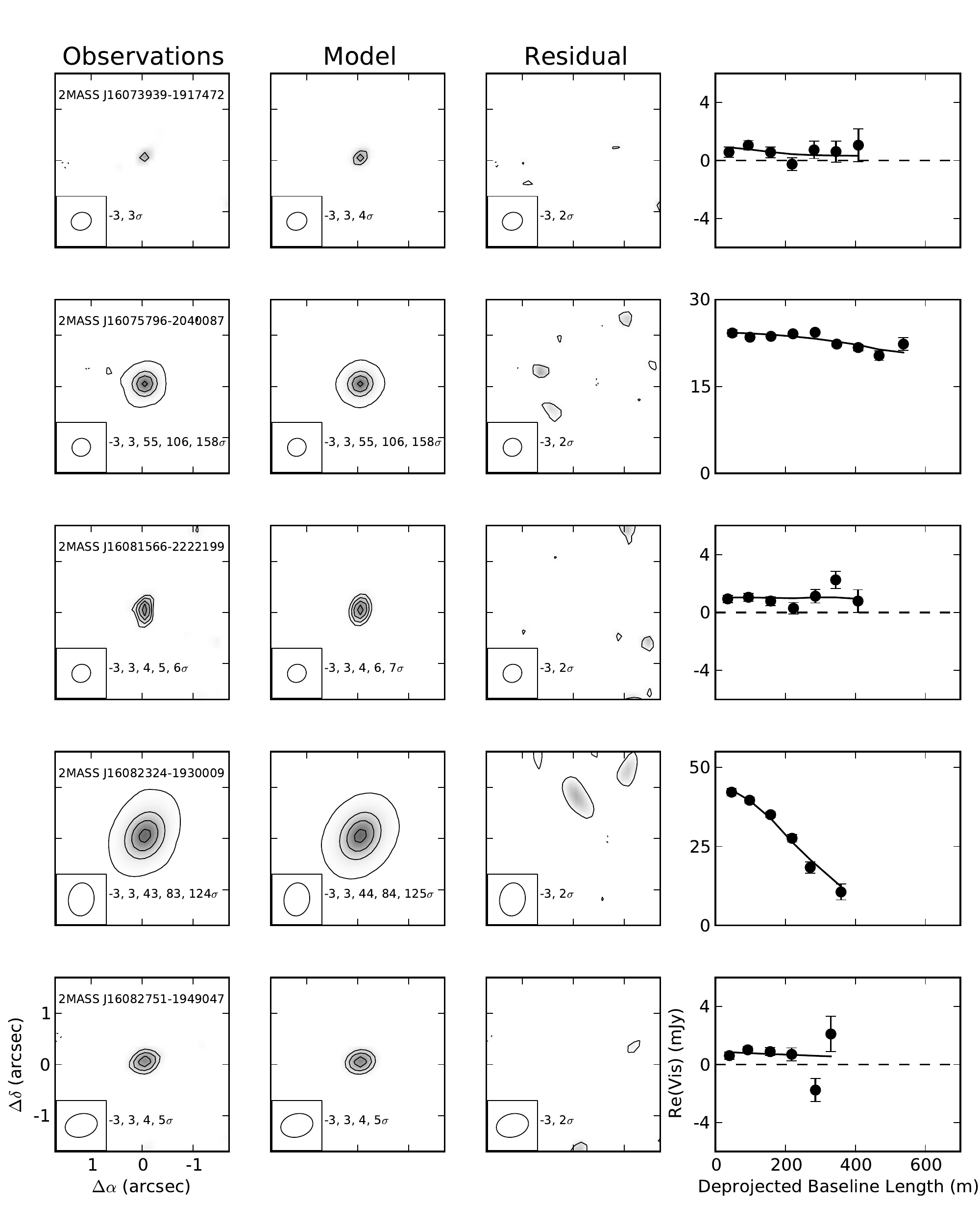}
}
\caption{Continued}
\label{fig:dust_results}
\end{figure}

\begin{figure}
\ContinuedFloat
\centering
\subfloat[][]{
\includegraphics[width=\textwidth]{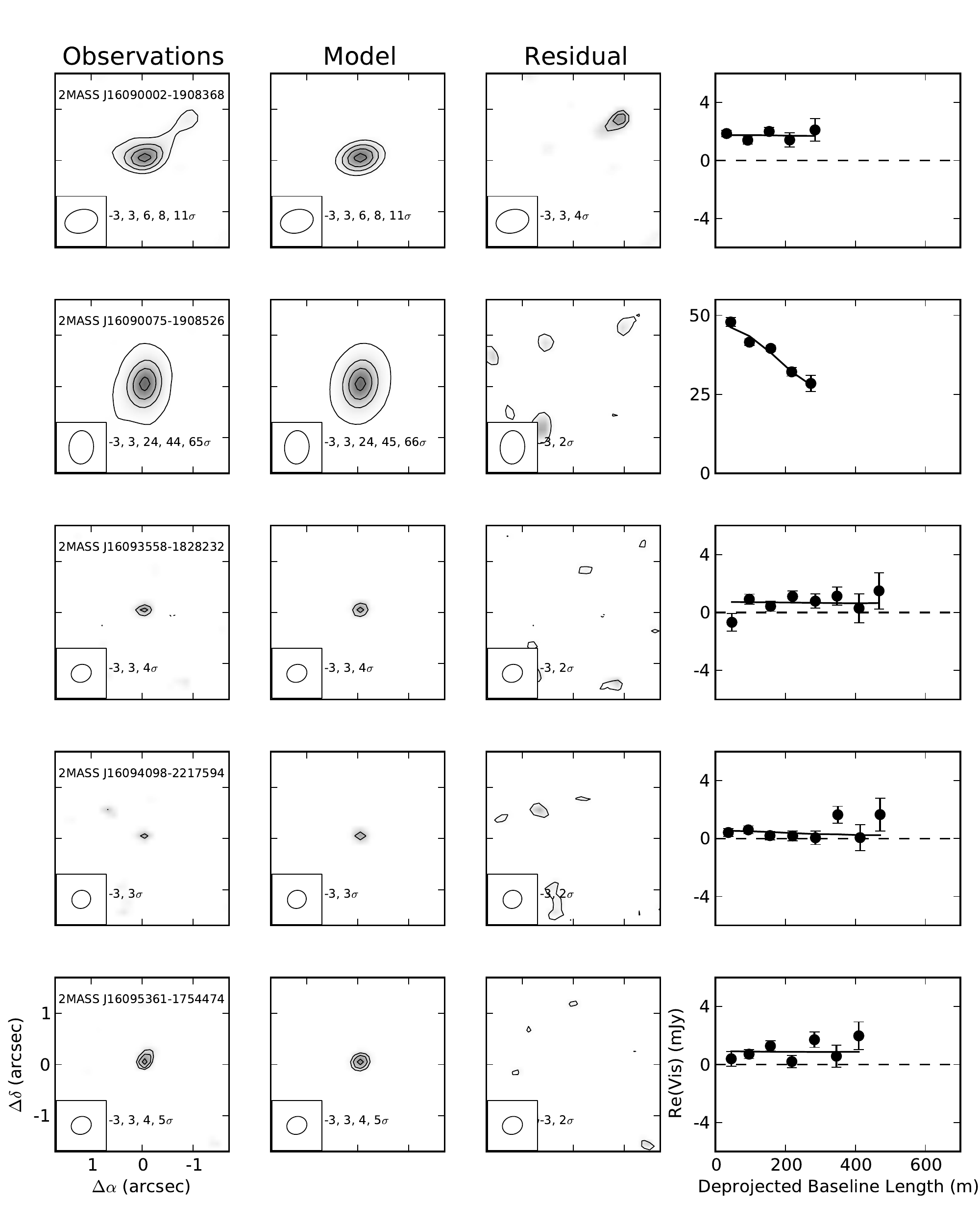}
}
\caption{Continued}
\label{fig:dust_results}
\end{figure}

\begin{figure}
\ContinuedFloat
\centering
\subfloat[][]{
\includegraphics[width=\textwidth]{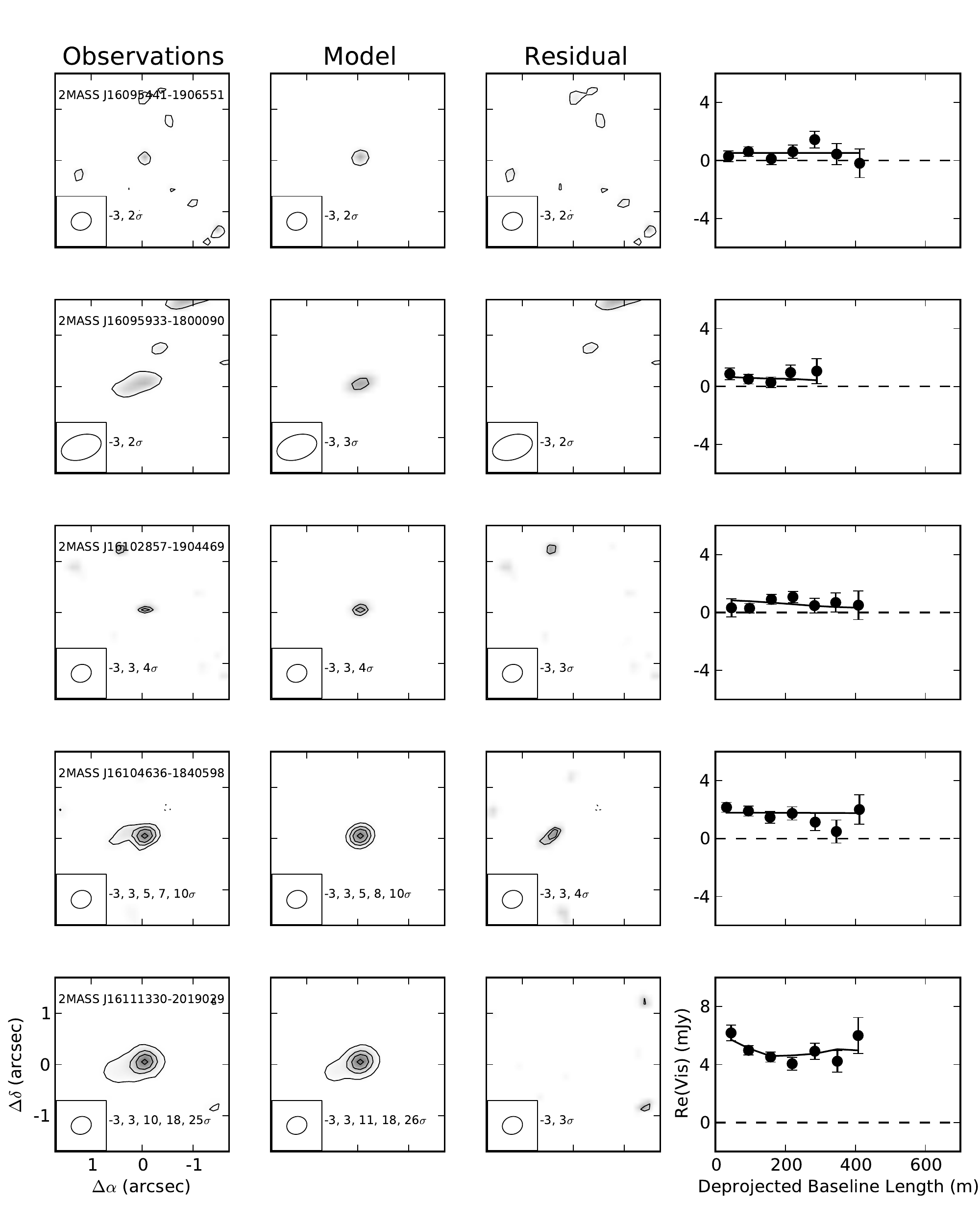}
}
\caption{Continued}
\label{fig:dust_results}
\end{figure}

\begin{figure}
\ContinuedFloat
\centering
\subfloat[][]{
\includegraphics[width=\textwidth]{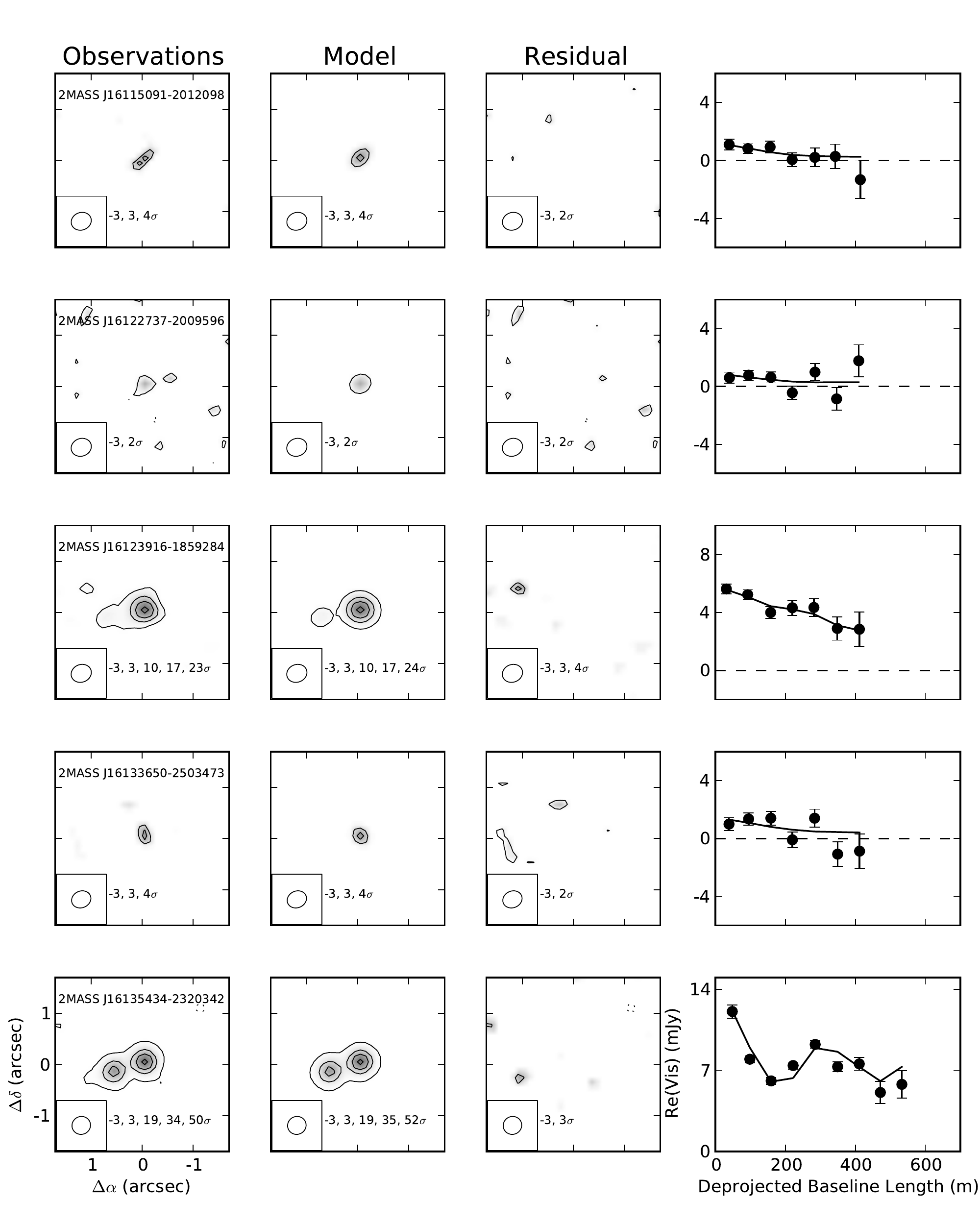}
}
\caption{Continued}
\label{fig:dust_results}
\end{figure}

\begin{figure}
\ContinuedFloat
\centering
\subfloat[][]{
\includegraphics[width=\textwidth]{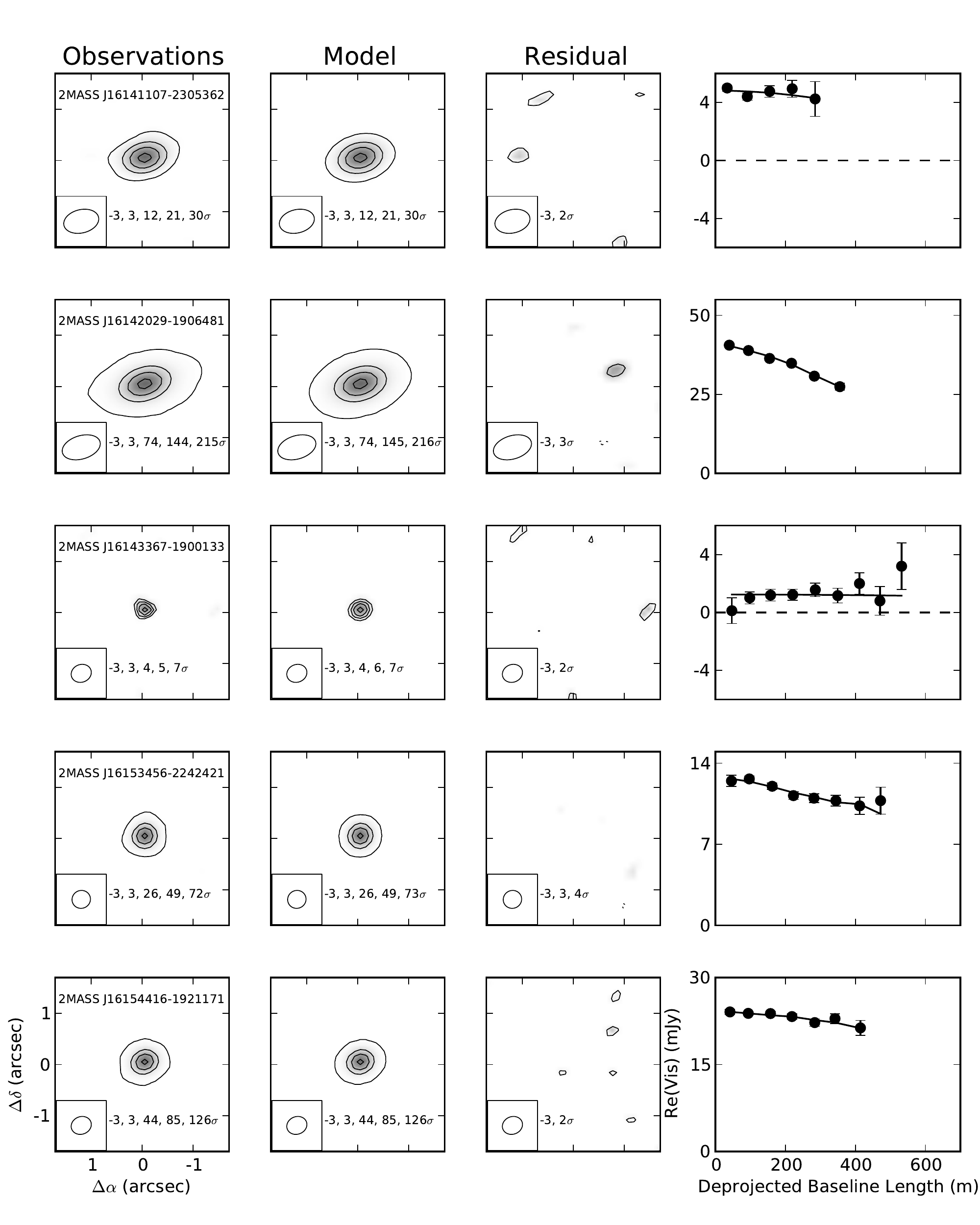}
}
\caption{Continued}
\label{fig:dust_results}
\end{figure}

\begin{figure}
\ContinuedFloat
\centering
\subfloat[][]{
\includegraphics[width=\textwidth]{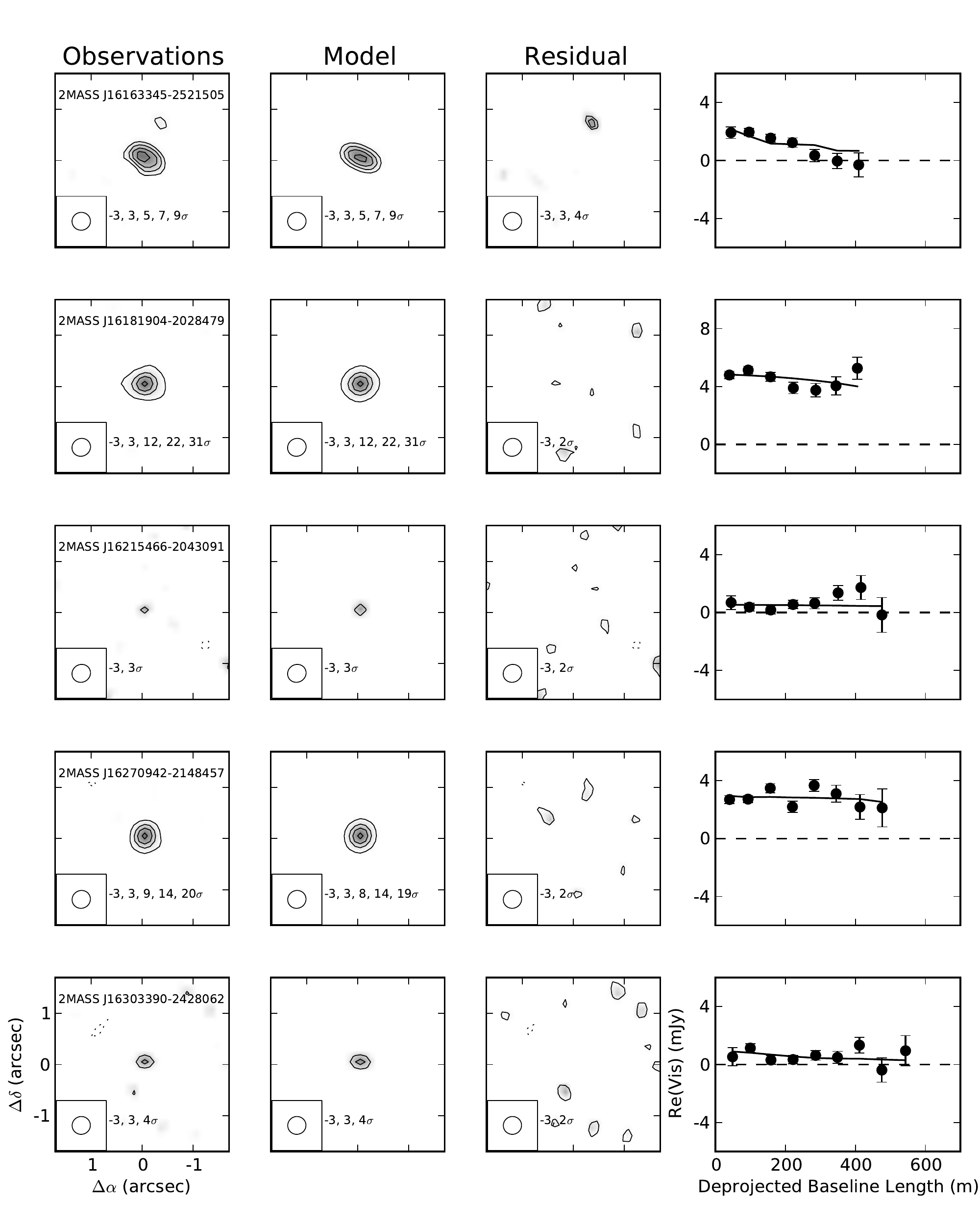}
}
\caption{Continued}
\label{fig:dust_results}
\end{figure}

\begin{figure}[!h]
\centering

\subfloat[][]{
\includegraphics[width=\textwidth]{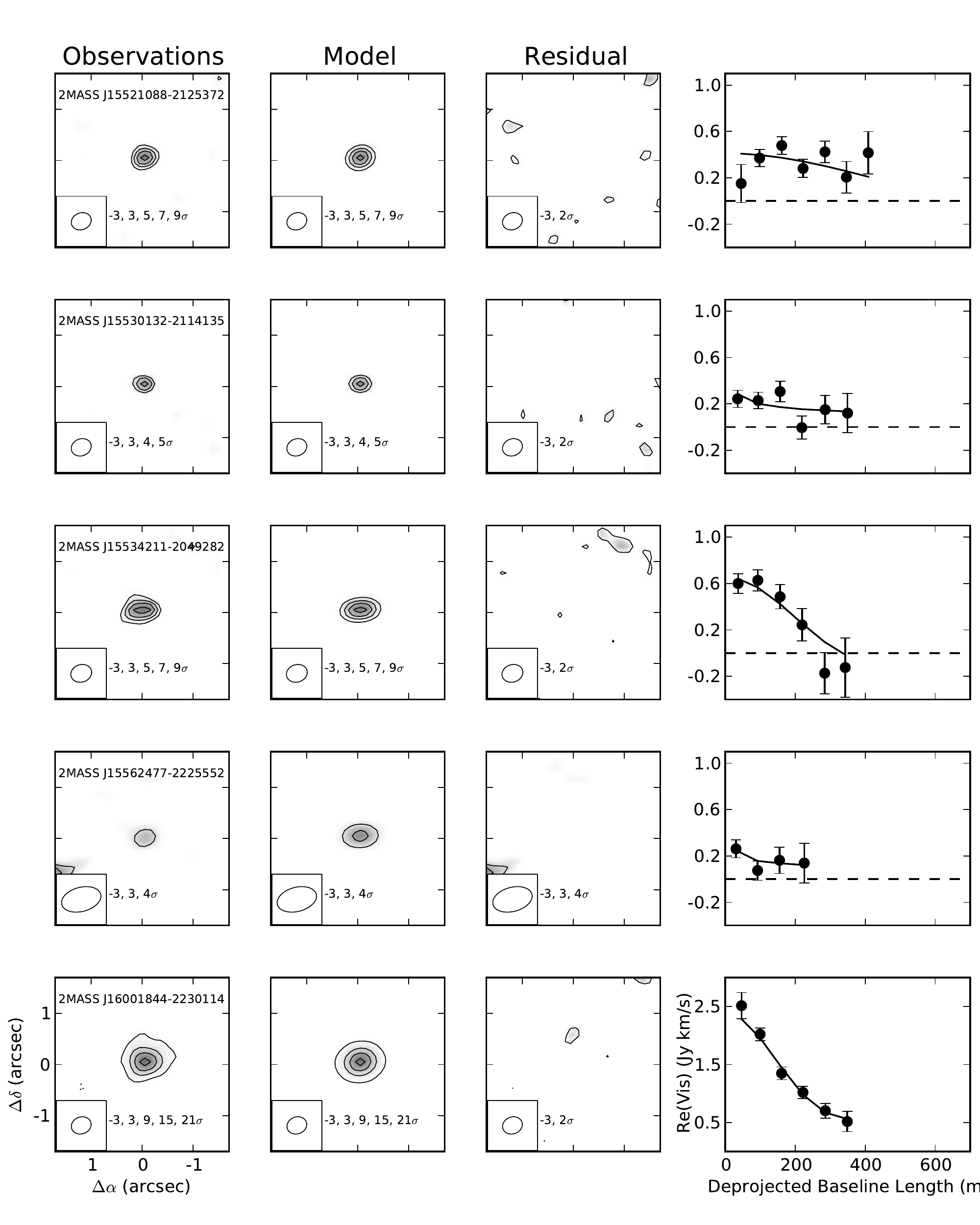}
}
\caption{Observed, model, and residual images corresponding to the best fit CO model parameters for each source.  The real part of the deprojected visibilities for the observations 
(solid points) and best fit model (solid curve) are also shown as a function of baseline length.}
\label{fig:CO_results}
\end{figure}

\begin{figure}
\ContinuedFloat
\centering
\subfloat[][]{
\includegraphics[width=\textwidth]{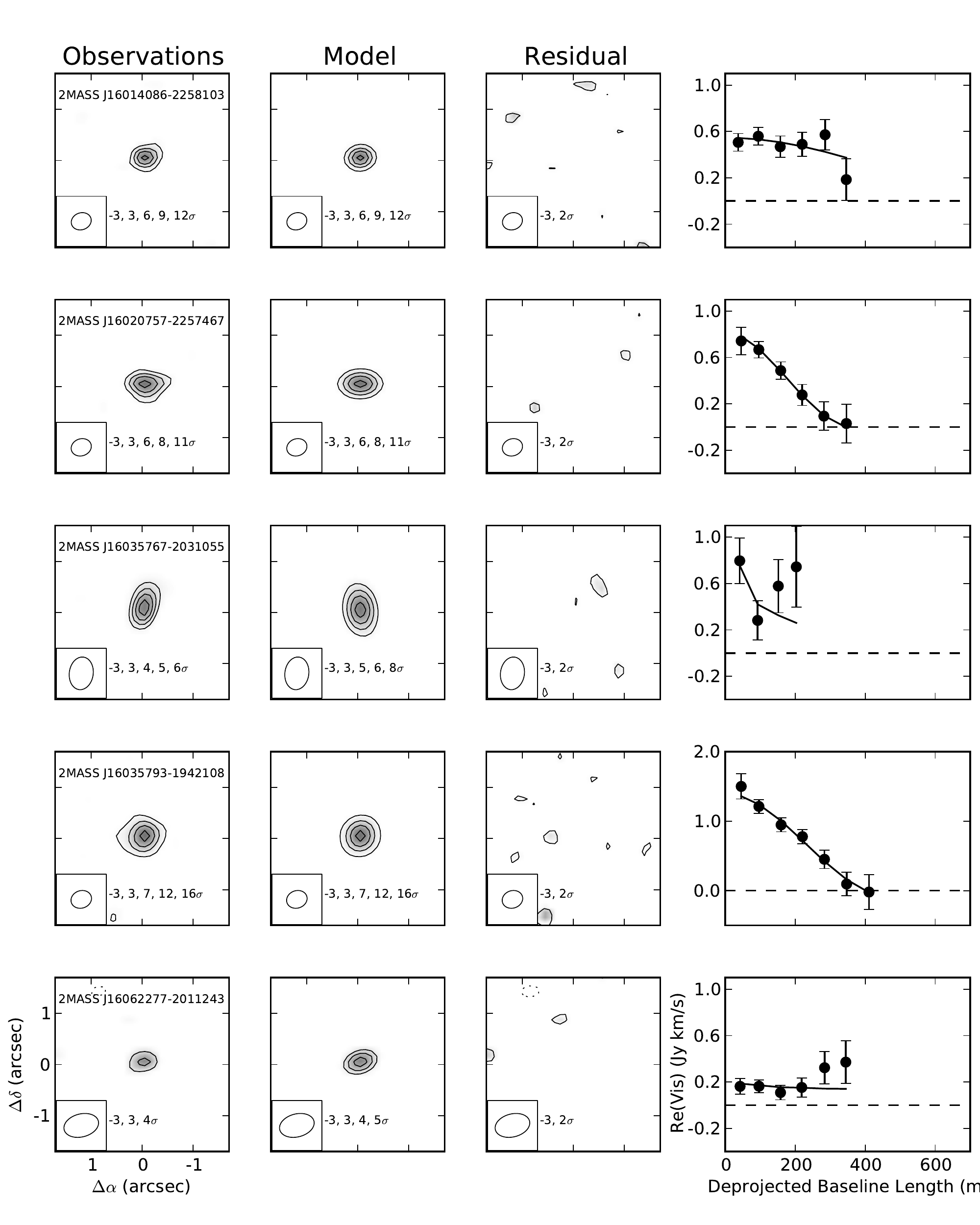}
}
\caption{Continued.}
\label{fig:CO_results}

\end{figure}

\begin{figure}
\ContinuedFloat
\centering
\subfloat[][]{
\includegraphics[width=\textwidth]{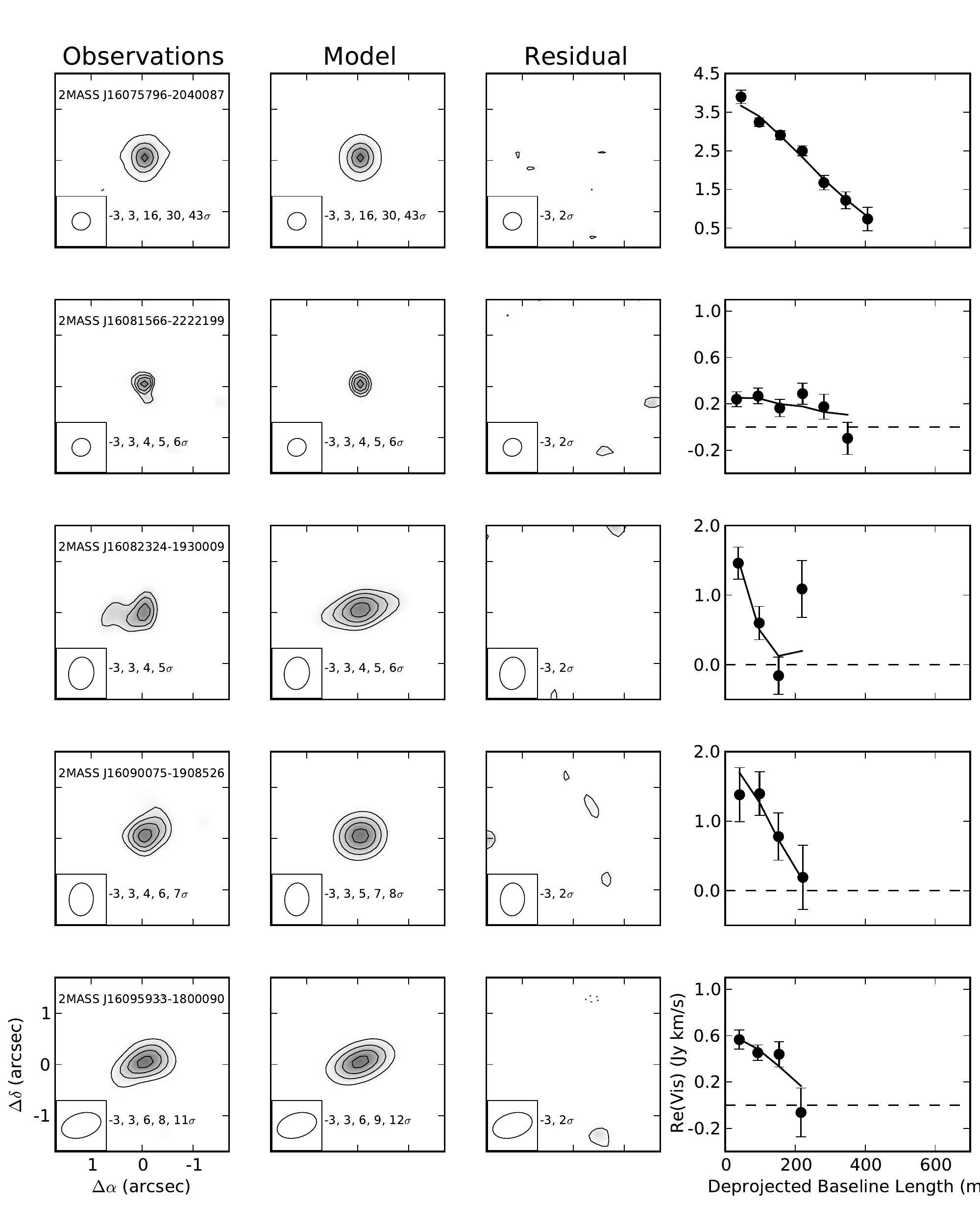}
}
\caption{Continued}
\label{fig:CO_results}
\end{figure}

\begin{figure}
\ContinuedFloat
\centering
\subfloat[][]{
\includegraphics[width=\textwidth]{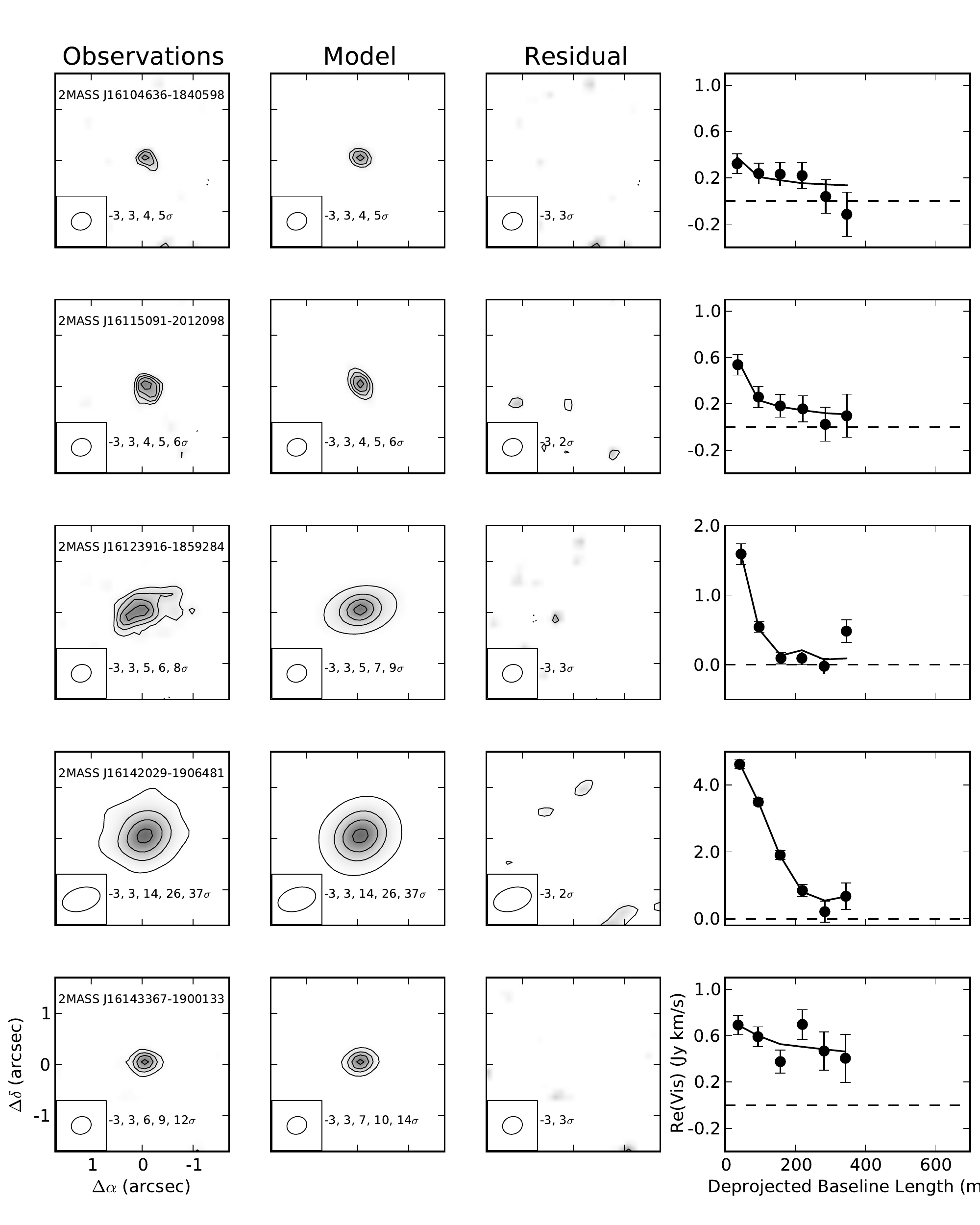}
}
\caption{Continued}
\label{fig:CO_results}
\end{figure}

\begin{figure}
\ContinuedFloat
\centering
\subfloat[][]{
\includegraphics[width=\textwidth]{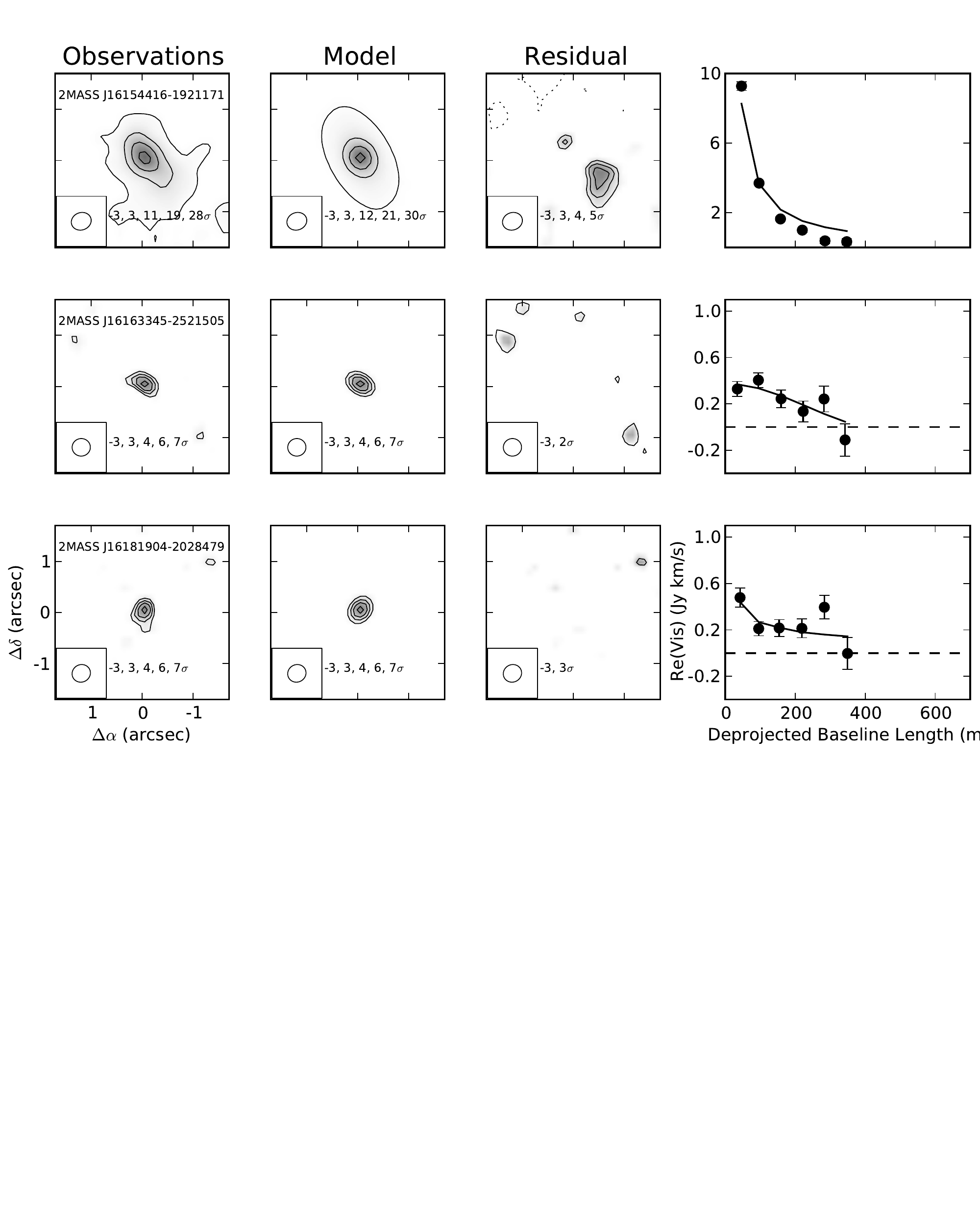}
}
\caption{Continued}
\label{fig:CO_results}
\end{figure}

\newpage

\begin{deluxetable}{lcccccc}
\tabletypesize{\tiny}
\tablecolumns{7} 
\tablewidth{0pt} 
\tablecaption{Continuum Fitting Results}
\tablehead{
  \colhead{Source}   &
  \colhead{$\log\frac{\Sigma_0}{\mathrm{g cm}^{-2}}$}  &
  \colhead{$R_{dust}$ (au)} &
  \colhead{$p$}  &
  \colhead{$h_0$ (au)}  &
  \colhead{Inclination (deg)}  &
  \colhead{Position Angle (deg)}
}
\startdata
2MASS J15354856-2958551  & -2.45 (-0.19,+0.24) & 14 (-10,+18)  &  1.12 (-0.04,+0.06) &  3 (-3,+7)  &  46 (-34,+17) &  25 (-25,+121)  \\
2MASS J15514032-2146103  & -2.78 (-0.25,+0.29) & 13 (-10,+35)  &  1.12 (-0.04,+0.06) &  8 (-8,+1)  &  84 (-59,+4)  &  130 (-95,+46)  \\
2MASS J15530132-2114135  & -1.41 (-0.20,+0.32) & 8 (-2,+5)     &  1.12 (-0.04,+0.05) &  8 (-4,+6)  &  47 (-40,+14) &  28 (-25,+117)  \\
2MASS J15534211-2049282  & -2.01 (-0.61,+1.07) & 45 (-7,+21)   &  1.13 (-0.06,+0.04) &  1 (-1,+1)  &  89 (-2,+1)   &  73 (-6,+5)     \\
2MASS J15582981-2310077  & -1.58 (-0.14,+0.21) & 13 (-3,+10)   &  1.13 (-0.07,+0.04) &  6 (-1,+1)  &  32 (-21,+18) &  47 (-32,+115)  \\
2MASS J16001844-2230114  & -1.98 (-0.13,+0.16) & 30 (-9,+5)    &  1.13 (-0.05,+0.05) &  5 (-3,+6)  &  45 (-21,+35) &  6 (-48,+36)    \\
2MASS J16014086-2258103  & -2.12 (-0.17,+0.14) & 36 (-9,+9)    &  1.12 (-0.06,+0.05) &  8 (-3,+10) &  74 (-31,+10) &  26 (-23,+22)   \\
2MASS J16014157-2111380  & -2.60 (-0.36,+0.38) & 9 (-9,+18)    &  1.13 (-0.05,+0.05) &  3 (-3,+6)  &  80 (-58,+5)  &  160 (-105,+20) \\
2MASS J16020757-2257467  & -2.17 (-0.12,+0.21) & 47 (-7,+8)    &  1.12 (-0.03,+0.07) &  1 (-1,+16) &  57 (-19,+14) &  80 (-15,+17)   \\
2MASS J16024152-2138245  & -1.37 (-0.17,+0.14) & 24 (-3,+3)    &  1.13 (-0.04,+0.05) &  8 (-3,+7)  &  41 (-21,+14) &  63 (-21,+28)   \\
2MASS J16030161-2207523  & -1.86 (-0.19,+0.24) & 19 (-8,+7)    &  1.13 (-0.04,+0.05) &  7 (-4,+8)  &  52 (-42,+22) &  62 (-50,+46)   \\
2MASS J16032225-2413111  & -2.20 (-0.19,+0.30) & 15 (-11,+13)  &  1.12 (-0.04,+0.04) &  6 (-3,+12) &  64 (-36,+16) &  72 (-43,+56)   \\
2MASS J16035767-2031055  & -2.51 (-0.10,+0.10) & 115 (-46,+88) &  1.11 (-0.05,+0.05) &  10 (-1,+1) &  69 (-27,+21) &  5 (-26,+22)    \\
2MASS J16035793-1942108  & -2.75 (-0.09,+0.10) & 173 (-60,+46) &  1.14 (-0.05,+0.05) &  9 (-1,+1)  &  56 (-34,+14) &  42 (-42,+34)   \\
2MASS J16041740-1942287  & -2.62 (-0.24,+0.34) & 9 (-8,+14)    &  1.12 (-0.05,+0.05) &  4 (-3,+6)  &  80 (-50,+7)  &  100 (-79,+60)  \\
2MASS J16043916-1942459  & -2.53 (-1.36,+1.10) & 46 (-42,+21)  &  1.13 (-0.05,+0.05) &  4 (-4,+7)  &  77 (-54,+9)  &  22 (-18,+123)  \\
2MASS J16052556-2035397  & -2.46 (-0.19,+0.28) & 16 (-12,+38)  &  1.21 (-0.05,+0.05) &  3 (-1,+1)  &  74 (-23,+16) &  91 (-68,+72)   \\
2MASS J16054540-2023088  & -1.72 (-0.08,+0.07) & 19 (-2,+5)    &  1.22 (-0.04,+0.06) &  9 (-1,+1)  &  67 (-29,+9)  &  10 (-10,+36)   \\
2MASS J16062196-1928445  & -2.77 (-0.13,+0.13) & 46 (-16,+32)  &  1.13 (-0.05,+0.05) &  13 (-1,+1) &  85 (-68,+5)  &  121 (-52,+39)  \\
2MASS J16062277-2011243  & -2.99 (-0.34,+0.33) & 9 (-8,+33)    &  1.22 (-0.05,+0.05) &  6 (-1,+1)  &  85 (-50,+5)  &  161 (-127,+13) \\
2MASS J16063539-2516510  & -2.28 (-0.20,+0.16) & 43 (-19,+17)  &  1.14 (-0.06,+0.04) &  8 (-3,+8)  &  74 (-43,+13) &  11 (-11,+70)   \\
2MASS J16064102-2455489  & -1.96 (-0.12,+0.24) & 29 (-8,+8)    &  1.13 (-0.05,+0.05) &  6 (-4,+8)  &  40 (-36,+14) &  81 (-41,+48)   \\
2MASS J16064385-1908056  & -3.04 (-0.20,+0.23) & 17 (-16,+62)  &  1.19 (-0.05,+0.05) &  3 (-1,+1)  &  48 (-39,+38) &  81 (-36,+81)   \\
2MASS J16072625-2432079  & -1.50 (-0.13,+0.20) & 29 (-2,+2)    &  1.13 (-0.05,+0.04) &  6 (-4,+4)  &  43 (-17,+10) &  2 (-14,+19)    \\
2MASS J16072747-2059442  & -2.19 (-0.17,+0.26) & 11 (-5,+9)    &  1.12 (-0.04,+0.06) &  4 (-4,+6)  &  68 (-49,+10) &  20 (-20,+106)  \\
2MASS J16073939-1917472  & -3.10 (-0.34,+0.33) & 9 (-9,+73)    &  1.13 (-0.05,+0.05) &  2 (-1,+8)  &  83 (-75,+7)  &  148 (-117,+31) \\
2MASS J16075796-2040087  & -0.64 (-0.22,+0.13) & 11 (-1,+1)    &  1.18 (-0.04,+0.04) &  18 (-4,+1) &  47 (-14,+8)  &  0 (-14,+15)    \\
2MASS J16081566-2222199  & -2.89 (-0.16,+0.19) & 80 (-41,+59)  &  1.13 (-0.05,+0.05) &  8 (-6,+5)  &  86 (-26,+4)  &  173 (-18,+24)  \\
2MASS J16082324-1930009  & -1.10 (-0.15,+0.18) & 65 (-5,+5)    &  1.16 (-0.07,+0.04) &  8 (-1,+1)  &  74 (-4,+5)   &  123 (-2,+3)    \\
2MASS J16082751-1949047  & -2.72 (-0.19,+0.21) & 44 (-35,+21)  &  1.11 (-0.05,+0.05) &  2 (-1,+1)  &  41 (-34,+34) &  17 (-11,+132)  \\
2MASS J16090002-1908368  & -2.57 (-0.14,+0.31) & 9 (-7,+18)    &  1.24 (-0.05,+0.06) &  19 (-1,+1) &  63 (-45,+18) &  84 (-38,+81)   \\
2MASS J16090075-1908526  & -1.27 (-0.06,+0.07) & 58 (-4,+5)    &  1.13 (-0.05,+0.05) &  6 (-1,+1)  &  56 (-5,+5)   &  149 (-9,+9)    \\
2MASS J16093558-1828232  & -2.87 (-0.31,+0.36) & 7 (-7,+28)    &  1.13 (-0.05,+0.05) &  3 (-3,+6)  &  83 (-59,+6)  &  104 (-81,+40)  \\
2MASS J16094098-2217594  & -3.58 (-0.35,+0.36) & 12 (-10,+62)  &  1.13 (-0.05,+0.05) &  1 (-1,+13) &  82 (-61,+6)  &  74 (-53,+65)   \\
2MASS J16095361-1754474  & -2.76 (-0.21,+0.31) & 6 (-6,+28)    &  1.18 (-0.05,+0.05) &  9 (-1,+1)  &  86 (-60,+4)  &  154 (-131,+16) \\
2MASS J16095441-1906551  & -3.11 (-0.58,+0.52) & 7 (-7,+41)    &  1.13 (-0.05,+0.05) &  2 (-2,+7)  &  83 (-72,+5)  &  177 (-42,+48)  \\
2MASS J16095933-1800090  & -3.56 (-0.30,+0.34) & 8 (-6,+63)    &  1.14 (-0.04,+0.06) &  16 (-1,+1) &  86 (-66,+4)  &  105 (-64,+59)  \\
2MASS J16102857-1904469  & -3.02 (-0.44,+0.40) & 9 (-9,+28)    &  1.14 (-0.06,+0.05) &  2 (-2,+12) &  84 (-51,+6)  &  98 (-74,+43)   \\
2MASS J16104636-1840598  & -2.13 (-0.25,+0.36) & 10 (-8,+15)   &  1.12 (-0.04,+0.06) &  8 (-5,+8)  &  71 (-63,+8)  &  84 (-38,+78)   \\
2MASS J16111330-2019029  & -1.69 (-0.27,+0.15) & 8 (-2,+8)     &  1.14 (-0.06,+0.04) &  6 (-3,+12) &  17 (-13,+40) &  141 (-78,+35)  \\
2MASS J16115091-2012098  & -2.94 (-0.25,+0.24) & 95 (-53,+6)   &  1.13 (-0.04,+0.06) &  1 (-1,+8)  &  86 (-42,+4)  &  144 (-44,+32)  \\
2MASS J16122737-2009596  & -2.98 (-0.30,+0.35) & 86 (-43,+15)  &  1.13 (-0.05,+0.05) &  1 (-1,+8)  &  26 (-14,+50) &  159 (-112,+18) \\
2MASS J16123916-1859284  & -2.21 (-0.10,+0.20) & 48 (-7,+8)    &  1.12 (-0.05,+0.05) &  8 (-5,+8)  &  51 (-36,+14) &  46 (-27,+22)   \\
2MASS J16133650-2503473  & -2.82 (-0.26,+0.26) & 45 (-33,+48)  &  1.14 (-0.06,+0.04) &  4 (-2,+12) &  86 (-52,+4)  &  23 (-29,+105)  \\
2MASS J16135434-2320342  & -1.18 (-0.59,+0.86) & 10 (-3,+3)    &  1.14 (-0.05,+0.05) &  6 (-5,+5)  &  52 (-44,+14) &  75 (-49,+52)   \\
2MASS J16135434-2320342B & -1.60 (-0.17,+0.25) & 13 (-3,+5)    &  1.14 (-0.05,+0.04) &  6 (-1,+12) &  40 (-34,+10) &  154 (-88,+29)  \\
2MASS J16141107-2305362  & -2.28 (-0.07,+0.12) & 30 (-8,+9)    &  1.04 (-0.04,+0.04) &  3 (-1,+1)  &  4 (-3,+48)   &  46 (-40,+104)  \\
2MASS J16142029-1906481  & -1.03 (-0.12,+0.17) & 29 (-2,+1)    &  1.10 (-0.02,+0.06) &  9 (-5,+1)  &  27 (-23,+10) &  19 (-19,+32)   \\
2MASS J16143367-1900133  & -2.69 (-0.18,+0.29) & 11 (-9,+13)   &  1.14 (-0.06,+0.05) &  8 (-6,+5)  &  69 (-43,+18) &  51 (-38,+109)  \\
2MASS J16153456-2242421  & -1.63 (-0.18,+0.11) & 21 (-2,+2)    &  1.12 (-0.04,+0.06) &  3 (-2,+16) &  46 (-21,+12) &  170 (-31,+10)  \\
2MASS J16154416-1921171  & -0.88 (-0.21,+0.25) & 10 (-1,+2)    &  1.15 (-0.05,+0.05) &  9 (-1,+8)  &  40 (-17,+24) &  117 (-54,+26)  \\
2MASS J16163345-2521505  & -2.33 (-0.57,+0.53) & 72 (-23,+25)  &  1.12 (-0.05,+0.06) &  1 (-1,+2)  &  88 (-9,+2)   &  64 (-9,+9)     \\
2MASS J16181904-2028479  & -1.62 (-0.18,+0.29) & 11 (-3,+6)    &  1.13 (-0.05,+0.06) &  8 (-5,+5)  &  56 (-46,+7)  &  90 (-56,+42)   \\
2MASS J16215466-2043091  & -3.08 (-0.55,+0.48) & 8 (-8,+29)    &  1.13 (-0.05,+0.05) &  1 (-1,+8)  &  82 (-53,+8)  &  127 (-110,+41) \\
2MASS J16270942-2148457  & -1.96 (-0.13,+0.25) & 22 (-6,+10)   &  1.13 (-0.05,+0.05) &  8 (-6,+7)  &  70 (-33,+15) &  176 (-29,+25)  \\
2MASS J16303390-2428062  & -2.98 (-0.23,+0.27) & 96 (-66,+3)   &  1.13 (-0.05,+0.05) &  1 (-1,+8)  &  74 (-25,+16) &  76 (-47,+75)   \\

\enddata
\label{tab:cont_fits}
\end{deluxetable}

\begin{deluxetable}{lccccc}
\tabletypesize{\tiny}
\tablecolumns{6} 
\tablewidth{0pt} 
\tablecaption{CO Fitting Results}
\tablehead{
  \colhead{Source}   &
  \colhead{$\log\frac{S_0}{Jy km/s/arcsecond^2}$}  &
  \colhead{$\gamma$}  &
  \colhead{$R_{CO}$ (au)} &
  \colhead{Inclination (deg)}  &
  \colhead{Position Angle (deg)}
}
\startdata
2MASS J15521088-2125372 & 0.77 (-0.18,+0.32) & 1.71 (-1.32,+0.05) &  24 (-13,+11)  &  24 (-17,+39) &  89 (-61,+57)  \\
2MASS J15530132-2114135 & 0.47 (-0.23,+0.51) & 1.70 (-0.82,+0.08) &  17 (-17,+37)  &  88 (-61,+2)  &  70 (-64,+47) \\
2MASS J15534211-2049282 & 0.96 (-0.21,+0.26) & 0.05 (-0.05,+0.63) &  51 (-10,+10)  &  77 (-10,+8)  &  95 (-13,+11)   \\
2MASS J15562477-2225552 & 0.26 (-0.39,+0.56) & 1.67 (-0.77,+0.12) &  16 (-16,+104) &  85 (-67,+5)  &  53 (-27,+79) \\
2MASS J16001844-2230114 & 1.20 (-0.03,+0.04) & 1.32 (-0.18,+0.11) &  68 (-14,+16)  &  24 (-10,+27) &  90 (-39,+41)  \\
2MASS J16014086-2258103 & 1.17 (-0.31,+0.53) & 1.71 (-1.19,+0.06) &  12 (-11,+13)  &  87 (-50,+3)  &  72 (-42,+55) \\
2MASS J16020757-2257467 & 0.77 (-0.09,+0.14) & 0.94 (-0.63,+0.26) &  54 (-11,+13)  &  59 (-18,+12) &  82 (-15,+16)  \\
2MASS J16035767-2031055 & 0.71 (-0.19,+0.33) & 1.69 (-0.51,+0.07) &  37 (-7,+182)  &  55 (-38,+23) &  22 (-95,+30) \\
2MASS J16035793-1942108 & 1.00 (-0.08,+0.06) & 0.79 (-0.58,+0.19) &  43 (-6,+7)    &  43 (-24,+10) &  3 (-28,+23) \\
2MASS J16062277-2011243 & 0.35 (-0.27,+0.78) & 1.70 (-1.33,+0.07) &  6 (-6,+37)    &  88 (-58,+2)  &  160 (-131,+12) \\
2MASS J16075796-2040087 & 1.67 (-0.04,+0.04) & 0.74 (-0.50,+0.31) &  34 (-2,+6)    &  52 (-5,+4)   &  1 (-5,+5) \\
2MASS J16081566-2222199 & 0.43 (-0.12,+0.18) & 1.67 (-0.34,+0.09) &  30 (-14,+129) &  3 (-2,+52)   &  15 (-88,+50) \\
2MASS J16082324-1930009 & 0.87 (-0.22,+0.34) & 0.95 (-0.38,+0.37) &  156 (-32,+29) &  72 (-11,+12) &  101 (-12,+14)  \\
2MASS J16090075-1908526 & 0.97 (-0.23,+0.17) & 1.30 (-0.91,+0.26) &  72 (-22,+42)  &  50 (-41,+10) &  95 (-40,+62)  \\
2MASS J16095933-1800090 & 0.65 (-0.13,+0.23) & 1.69 (-0.96,+0.08) &  52 (-23,+33)  &  63 (-42,+16) &  119 (-42,+40)  \\
2MASS J16104636-1840598 & 0.45 (-0.16,+0.36) & 1.70 (-0.40,+0.08) &  22 (-22,+148) &  54 (-33,+25) &  50 (-43,+68) \\
2MASS J16115091-2012098 & 0.54 (-0.14,+0.56) & 1.30 (-0.26,+0.26) &  75 (-29,+241) &  86 (-36,+4)  &  26 (-13,+13) \\
2MASS J16123916-1859284 & 0.70 (-0.07,+0.07) & 0.85 (-0.13,+0.09) &  169 (-26,+24) &  53 (-8,+6)   &  104 (-11,+14)  \\
2MASS J16142029-1906481 & 1.52 (-0.05,+0.07) & 0.97 (-0.12,+0.09) &  88 (-6,+6)    &  58 (-4,+4)   &  5 (-4,+4)  \\
2MASS J16143367-1900133 & 1.32 (-0.36,+0.35) & 1.71 (-0.96,+0.07) &  14 (-12,+12)  &  83 (-51,+5)  &  88 (-50,+76)  \\
2MASS J16154416-1921171 & 1.73 (-0.01,+0.01) & 1.01 (-0.01,+0.01) &  430 (-10,+10) &  61 (-1,+1)   &  28 (-1,+1)   \\
2MASS J16163345-2521505 & 0.79 (-0.23,+0.36) & 1.31 (-0.90,+0.27) &  45 (-19,+22)  &  81 (-17,+7)  &  59 (-19,+16) \\
2MASS J16181904-2028479 & 0.48 (-0.11,+0.28) & 1.69 (-0.26,+0.07) &  26 (-11,+176) &  69 (-60,+5)  &  155 (-28,+59)  \\
\enddata
\label{tab:CO_fits}
\end{deluxetable}

\begin{deluxetable}{lccccc}
\tabletypesize{\footnotesize}
\tablecolumns{6} 
\tablewidth{0pt} 
\tablecaption{Secondary Source Properties}
\tablehead{
  \colhead{Field}   &
  \colhead{$S_{tot}$ (mJy)} & 
  \colhead{$\Delta\alpha$ (arcsec)} & 
  \colhead{$\Delta\delta$ (arcsec)} 
}
\startdata
2MASS J16032225-2413111  &  0.85 $\pm$ 0.14 & 0.80 $\pm$ 0.04 & 0.06 $\pm$ 0.04 \\
2MASS J16054540-2023088  &  1.00 $\pm$ 0.15 & 0.39 $\pm$ 0.04 & -0.01 $\pm$ 0.04 \\
2MASS J16111330-2019029  &  1.00 $\pm$ 0.16 & 0.48 $\pm$ 0.04 & -0.19 $\pm$ 0.04 \\
2MASS J16123916-1859284  &  1.09 $\pm$ 0.16 & 0.75 $\pm$ 0.04 & -0.15 $\pm$ 0.04 \\
2MASS J16135434-2320342  &  5.83 $\pm$ 0.12 & 0.59 $\pm$ 0.03 & -0.18 $\pm$ 0.03 \\
\enddata
\label{tab:secondaries}
\end{deluxetable}

\end{document}